\documentclass[12pt, a4 paper]{article}
\usepackage[paper=A4,pagesize]{typearea}
\usepackage[left=2.54cm, right=2.54cm, top=2.54cm, bottom=2.54cm]{geometry}
\usepackage[mathscr]{euscript}
\usepackage{graphicx}
\usepackage{amsmath}
\usepackage{newtxtext}
\usepackage{newtxmath}
\usepackage{natbib}
\usepackage{epstopdf,epsfig}
\usepackage{hyperref}
\hypersetup{
    colorlinks = true,
    urlcolor   = black,
    citecolor  = black,
}

\usepackage{multirow}
\usepackage{caption} 
\usepackage{float}
\usepackage{placeins}
\usepackage{titlesec}
\usepackage{bm}
\usepackage{enumerate}
\usepackage{capt-of}
\usepackage{booktabs}
\usepackage{cleveref}
\usepackage{array}
\newcommand{\PreserveBackslash}[1]{\let\temp=\\#1\let\\=\temp}
\newcolumntype{C}[1]{>{\PreserveBackslash\centering}p{#1}}
\newcolumntype{R}[1]{>{\PreserveBackslash\raggedleft}p{#1}}
\newcolumntype{L}[1]{>{\PreserveBackslash\raggedright}p{#1}}

\usepackage{caption} 
\usepackage{soul}
\usepackage[normalem]{ulem}
\usepackage{enumitem}

\newcommand\blfootnote[1]{%
	\begingroup
	\renewcommand\thefootnote{}\footnote{#1}%
	\addtocounter{footnote}{-1}%
	\endgroup
}

\usepackage{fancyhdr}
\usepackage{lineno}

\begin{document}
\title{Learning thermoacoustic interactions in combustors using a physics-informed neural network}

\author{Sathesh Mariappan\textsuperscript{a} \and Kamaljyoti Nath \textsuperscript{b} \and George Em Karniadakis \textsuperscript{b, c}}


\newcommand{\Addresses}{{%
			\textsuperscript{a}  Department of Aerospace Engineering, Indian Institute of Technology Kanpur, India\\
			\textsuperscript{b}  Division of Applied Mathematics, Brown University, USA\\
			\textsuperscript{c} School of Engineering, Brown University, USA
	}}
 \maketitle

 \vspace*{-0.5cm}
	\begin{center}
		\Addresses
	\end{center}
\blfootnote{
 \textit{E-mail addresses:} sathesh@iitk.ac.in (Sathesh Mariappan), 
 kamaljyoti\_nath@brown.edu (Kamaljyoti Nath), george\_karniadakis@brown.edu (George Em Karniadakis) }
	\thispagestyle{fancy}
	\cfoot{}
	\rhead{}
	\renewcommand{\headrulewidth}{0pt}
	\rfoot{\today}

	\date{}

\begin{abstract}
We introduce a physics-informed neural network (PINN) method to study thermoacoustic interactions leading to combustion instability in combustors. Specifically, we employ a PINN to investigate thermoacoustic interactions in a bluff body anchored flame combustor, representative of ramjet and industrial combustors. Vortex shedding and acoustic oscillations appear in such combustors, and their interactions lead to the phenomenon of vortex-acoustic lock-in. Acoustic pressure fluctuations at three locations and the total flame heat release rate serve as the measured data. The coupled parameterized model is based on the acoustic equations and the van der Pol oscillator for vortex shedding. The PINN was applied in the combustor, where the measurements suitable for a future machine learning application were not anticipated at the time of the experiments, as is the case in the vast majority of available data in the literature. We demonstrate a good performance of PINN in generating the acoustic field (pressure and velocity fluctuations) in the entire spatiotemporal domain, along with estimating all the parameters of the model. Therefore, this PINN-based model can potentially serve as an effective tool in improving existing combustors or designing new thermoacoustically stable and structurally efficient combustors. 

\end{abstract}

\section{Introduction}
\label{sec:Introduction}
Combustion-driven oscillation is a significant concern in land and aero-derivative gas turbine engines. Requirements of stricter pollution emission norms and increased power/thrust to weight ratio have led to the development of fuel-lean premixed pre-vaporized combustors. Such combustors are observed to show unwanted large-amplitude catastrophic oscillations, popularly known as combustion instability \citep{lieuwen2003combustion}. It appears as an even more significant challenge than before in land-based gas turbine engines, employing alternate fuel blends, such as Hydrogen \citep{taamallah2015fuel,strollo2021effect}, Ammonia \citep{valera2019premixed} etc., and staged combustors \citep{ma2023experimental}.
The instability occurs due to a positive feedback mechanism between the combustor acoustic field and the unsteady heat released by the flame. 

One of the challenging aspects of predicting the occurrence of combustion instability at the design stage is the lack of quantitatively accurate models. Large-eddy simulations (LES), although a possible solution, are expensive to perform for the entire parameter space of operation. A potentially effective approach is to use low-order models (LOMs) that capture the essential physics with tunable parameters. LOMs would reduce the parameter space for detailed simulations (such as LES) or experiments to identify the operation regime. With the rapid developments in machine learning, there have been significant recent efforts (\citealt{hachijo2019early,mondal2021transfer,juniper2023machine} to mention a few) to predict combustion instability based on purely data-driven and hybrid data+physics-based methods. Among the two, the latter option is more promising for the following reason. Though LOMs retain the essential physics, they do not model all the details of a practical combustor. The large set of available experimental data from the past, combined with the LOMs, allows the development of hybrid models. Such models are expected to be quantitatively more accurate than the LOMs. We propose to use the recently developed physics-informed neural network (PINN) \citep{raissi2019physics} to develop a hybrid neural network model for a thermoacoustic interaction, namely the vortex-acoustic lock-in.

In PINN, the unknown quantities, such as acoustic pressure and velocity (outputs), are approximated using neural networks, whose inputs are the independent variables (time and space coordinates). Weights and biases of the network are trainable parameters. In the case of inverse problems, LOM parameters are estimated simultaneously along with the dynamics of the system. PINN minimizes a total loss function consisting of physics and data losses. The minimization process trains the neural network and the parameters. The physics loss is obtained from the residue of the equations in the LOM, while the data loss is formed from the difference between the available experimental values of the dependent variables and the neural network output. Before we proceed further, we present a brief review of other physics-based machine learning methods employed in thermoacoustics. 

\subsection{Contemporary machine learning methods employed in thermoacoustics}
The authors in \citealt{garita2021assimilation,juniper2022generating,juniper2023machine} considered a physics-based Bayesian inference model in thermoacoustics. Their approach was the first to identify a set of candidate models containing tunable parameters. The goal was to identify the best model, its parameters, and their (Bayesian) uncertainties that best fit the experimental data. The probability of the parameters (prior) was assumed as given. The required inputs were data of relevance, such as growth rate, frequency, etc. They required measurements obtained from several experimental realizations for an operating condition. Experiments revealed the likelihood of the parameters. Combining the above two and using Bayes' theorem, an expression for the conditional probability of the parameters, given the experimental data and a model (posterior), was determined. This conditional probability was maximized with respect to the parameters to obtain the maximum likelihood of the parameters. The uncertainty associated with the parameters was determined from a Laplace method \citep{juniper2022generating}. The numerical value of the product of the likelihood and the prior integrated over the parameters, termed marginal likelihood, were used for choosing the best model that fitted the data. Other technical issues in the above procedure are described in \cite{magri2016stability}. The method was successfully applied to a Rijke tube with heated wires as the heat source to generate a quantitatively accurate thermoacoustic model \citep{garita2021assimilation,juniper2022generating}. The procedure required many experimental realizations ($\sim 10^3$). The simplicity of the setup allowed it to be automated for repeated realizations, which may not have been possible in many past experiments (for example \citealt{komarek10,dawson14,singh2021experimental}). 
Furthermore, \cite{juniper2022generating} suggested that the sensor placement has to be performed carefully, presuming the application of Bayesian inference. Many past experimental data do not have this presumption. In our work, we attempt to use PINN in such an experiment, where a machine learning method that would be applied was not presumed.  

Another physics-based machine learning method is the physics-informed echo state network (PI-ESN), developed by \cite{doan2020physics,doan2021short}. The neural network consisted of an input layer, which receives the dynamical state of the system at a given time step. The input layer was followed by a high-dimensional reservoir whose weights were randomly initialized. Neither the weights of the input layer nor the reservoir were trained. The reservoir was connected to an output layer to produce the dynamical state at the next time step. The difference between the output and the state of the system (known by other means such as numerical simulations and experiments) at the next time step formed one part of the loss function. Physics-based equations were introduced additionally in the loss function. The total loss function was minimized to obtain the weights of the output layer. The output layer was trained in a given time window. The trained PI-ESN forecasted long-time statistics of chaotic system \citep{doan2020physics} and turbulence \citep{doan2021short}. PI-ESN may be computationally cheap compared to PINN, as only the last layer of the neural network needs to be trained in the former. However, physics involving partial differential equations, inverse problems, and the determination of hidden variables, as in PINN, are yet to be explored. Additionally, to the best of the author's knowledge, PI-ESN is yet to be applied to thermoacoustic systems.

Other applications of neural networks include the determination of flame transfer function \citep{jaensch2017uncertainty,wu2023reconstruction}, precursors \citep{gangopadhyay2018characterizing,cellier2021detection,mccartney2022reducing}, control \citep{zhang2022neural} and acceleration of the numerical simulations \citep{shadram2022physics} of combustion instability. They are all purely data-driven.

\subsection{PINN selection in the current work}
PINN has the following merits compared to the existing physics-based machine learning methods. 1) It can be applied to experimental data measured at sparse spatial locations in the literature on thermoacoustics, where the presumption of its future use for machine learning was unknown. 2) It allows the identification of hidden states, which are otherwise challenging to measure, especially in practical systems. 3) The hybrid solution provided in the form of a neural network allows one to perform many mathematical operations (such as differentiation through automatic-differentiation described in \citealt{baydin2018automatic}) that one does with any analytical or numerical solutions. 4) Mathematically ill-posed problems can be solved by leveraging them, using the available experimental/numerical data at sparse locations. 5) The inverse problem of LOM parameter identification is still possible for such ill-posed problems. On the whole, new physical insights and designs can emerge from many of the past thermoacoustic experiments and their corresponding numerical simulations/LOMs by revisiting using PINN. 

PINN does come with some drawbacks. 1) Training of a relatively large number of neurons to attain a global minimum of the loss function is challenging. A careful landscaping of the loss function by altering the penalization weights of the individual loss terms is required. Algorithms such as self-adaptive weights \citep{mcclenny2020self} help alleviate the problem. 2) PINN involves tuning many hyperparameters (such as network size and learning rates), which lack a prescription to get the best solution. Therefore, the user needs to tune them by performing several runs. 3) Accurate long-time prediction is still a challenging task.  

Due to PINN's merits, since its first paper \citep{raissi2019physics}, an explosion of its improved versions such as conservative PINN (c-PINN) for conversation laws and domain-decomposition \citep{jagtap2020conservative}, Parareal PINN (PPINN, \citealt{Meng_2020}), backward compatible (bc-PINN \citealt{mattey2022novel}), sequence to sequence PINN \citep{krishnapriyan2021characterizing} for long time integration, variational formulation of PINN with domain decomposition (hp-VPINN) \citep{Kharazmi2021}, extended PINN (XPINN). Specifically, an improved version of c-PINN having parallelizable capability on large domains \citep{Jagtap_2020_XPINN}, separable PINN (sPINN, \citealt{cho_2022_separable}) to overcome the curse of dimensionality for higher dimensional problems have been developed. Since we apply PINN for the first time in thermoacoustics, we begin with the basic PINN \citep{raissi2019physics}, combined with self-adaptive weights \citep{mcclenny2020self} despite the above recent developments.

PINN has been successfully applied in a wide range of disciplines and engineering applications: soil mechanics \citep{Depina_2022_unsaturated}, bio-medicine flow \citep{arzani2021uncovering}, photonics \citep{Chen_2020_neno_optics}, diesel engines \citep{Nath_2023_Physics}, multistage centrifugal pumps \citep{carvalho2023learning}, to mention a few. Interested readers may refer to review articles by \cite{karniadakis_2021_PINN_review,Cola_2022_Scientific,Karami_2022_Review}.

\subsection{Employing PINN to study vortex-acoustic lock-in}
We choose to demonstrate PINN in a nonlinear thermoacoustic problem: lock-in. It is commonly observed in combustors, where the flame is anchored by a backward-facing step \citep{zukoski1985combustion,schadow92,chakravarthy07} and bluff body \citep{emerson2012frequency,emerson2015dynamics}. Such combustors are employed in many industrial heating systems and ramjet engines. A nonlinear feedback interaction between vortex shedding due to the flame anchoring methods and the combustor acoustic field leads to thermoacoustic oscillations occurring at a common frequency, known as the lock-in frequency. The phenomenon is termed as (frequency) lock-in. Generally, during lock-in, large amplitude combustion instability is reported \citep{zukoski1985combustion,schadow92,wee2004self,altay2009flame}; however, stable combustor operation is also observed in some cases \citep{emerson2015dynamics,guan2019open}.

The rest of the paper is arranged as follows. Section \ref{sec:combustor_description} provides a brief description of the combustor. It is followed by presenting the LOM and relevant experimental results in \S \ref{sec:lom}. We then describe the PINN framework and the optimization procedure in \S \ref{sec:PINN_arch}. Sections \ref{sec:sol_acoustic_eq} - \ref{sec:inv_prob_syn} demonstrate the applicability of PINN using simulated data. The next section uses experimental data, and accompanying results are elaborated (\S \ref{sec:inv_prob_exp}). The last section \ref{sec:conc} summarizes the findings.       

\section{Brief description of the combustor test rig}
\label{sec:combustor_description}
\begin{figure}
\centerline{\includegraphics[keepaspectratio,width=0.6\textwidth]{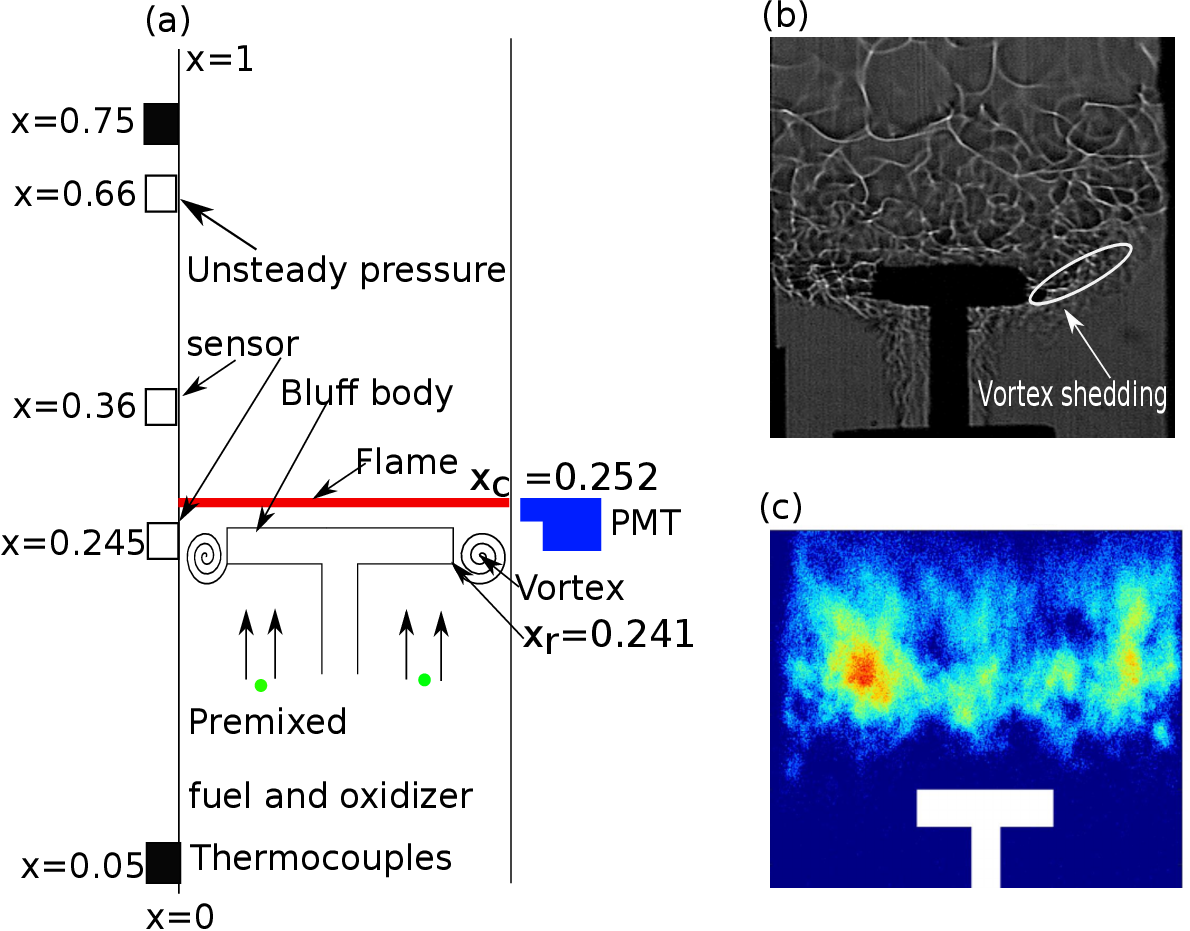}}
\caption{\textbf{Experimental configuration and measurement:} $(a)$ Schematic (not to scale) representation of the bluff body stabilized combustor test rig. The co-ordinates of various elements are indicated in their non-dimensional version. Hollow and filled rectangles indicate the acoustic sensor and thermocouple, respectively. The photomultiplier tube (PMT) is shown as the blue-shaded equipment. An instantaneous snapshot of the measured schlieren $(b)$ and CH\textsuperscript{*} chemiluminescence $(c)$ images, showing vortex shedding and flame, respectively. Red and blue colors indicate the maximum and minimum values in panel $(c)$.}
\label{fig:geometry_PINN_setup}
\end{figure}

The combustor consists of a 3 kW bluff body (circular disc) stabilized premixed liquefied petroleum gas-air burner (schematic shown in figure \ref{fig:geometry_PINN_setup}$a$). This burner is housed in a Rijke tube-type combustor of length $\tilde l=1$ m and square cross-section $\tilde A=65$ mm\textsuperscript{2}, where `` $\tilde ~$ " indicates a dimensional quantity. The bluff body is circular, having a diameter and thickness of 14 mm and 4 mm, respectively. The top of the (circular) bluff body is at an axial location of 0.245 m from the upstream end of the combustor. Apart from anchoring the flame, the bluff body sheds vortices from its bottom surface ($0.241$ m). An instantaneous schlieren snapshot is shown in figure \ref{fig:geometry_PINN_setup}$(b)$. The white circle shows the shedding and the path of the shed vortices. The flame sits slightly above the top surface of the bluff body. An instantaneous CH\textsuperscript{*} chemiluminescence image is indicated in figure \ref{fig:geometry_PINN_setup}$(c)$, where red and blue regions represent high and low levels of heat release rate. The vortices strongly perturb the downstream located flame at its shedding frequency. On the other hand, the acoustic field of the combustor poses another frequency. Unsteady acoustic pressure fluctuations at three axial locations, 0.245 m, 0.36 m, and 0.66 m (from the upstream end of the combustor), and the total CH\textsuperscript{*} flame chemiluminescence emission are the measurements (figure \ref{fig:geometry_PINN_setup}$a$). The airflow rate is the only operational parameter that is varied; it alters the vortex shedding frequency while not influencing the acoustic frequency (dictated by the length of the combustor tube). 

For a range of flow rates, lock-in is observed. We noticed a common frequency (and its harmonics) during the lock-in. The common frequency is either close to the vortex shedding or the acoustic frequency, accordingly termed V- and A-lock-in, respectively. A complete investigation of the experiments is discussed in \cite{singh2021experimental}, while its re-interpretation with a qualitatively accurate LOM for the two lock-in regions is in \cite{britto2023self}. The latter paper concluded that most of the A- and V-lock-in regions are associated with high (combustion instability) and low-amplitude acoustic pressure fluctuations, respectively.

\section{Dynamic low-order model for the combustor}
\label{sec:lom}
The previous section discussed the combustor configuration, followed by an introduction to the A-/V-lock-in regions. LOMs of the combustor acoustic field and flame dynamics due to vortex shedding are described in the first two subsections. The last subsection characterizes the salient features of the experimental observations. 

\subsection{Acoustic model of the combustor}
The idealized combustor configuration for the LOM is shown in figure \ref{fig:geometry_PINN_setup}$(a)$. In the experiments, we observe only longitudinal acoustic modes. Therefore, we use the one-dimensional acoustic wave equation (spatial $\tilde x$ - along the flow direction, temporal $\tilde t$ coordinates). The following set of equations represents the evolution of the dimensional acoustic pressure ($\tilde p(x,t)$) and velocity ($\tilde u(x,t)$) in the combustor. The acoustic equations are valid at low steady-state Mach numbers $(M_s)$, as in the experiments $M_s\sim 10^{-3}$.

\begin{align}
\frac{\partial \tilde u}{\partial \tilde t}+\tilde T_s\frac{\partial \tilde p}{\partial \tilde x}&=0\label{eq:dim_damped_wave_eq_temp_jump_mom},\\
       \frac{\partial \tilde p}{\partial \tilde t}+\frac{\partial \tilde u}{\partial \tilde x}+\tilde \alpha \tilde p&=\frac{(\gamma_s -1) \tilde q}{\tilde A}\frac{e^{-\left(\left(\tilde x - \tilde x_c\right)/\tilde b\right)^2}}{\tilde b\sqrt{\pi}}\label{eq:dim_damped_wave_eq_temp_jump_en}.
\end{align}

The parameters $\tilde \alpha,~\tilde A$ and $\gamma_s$ indicate the acoustic damping coefficient, cross-section area of the combustor, and the ratio of specific heat capacities of air, respectively. Among the parameters, $\tilde A$ and $\gamma_s=1.4$ for air are known. Acoustic damping is caused by the radiation of the waves at the ends and boundary layer formed inside of the combustor. It is true that representing it by a single parameter $\tilde \alpha$ may be inadequate to represent the true nature of the damping process \citep{lieuwen12,juniper2022generating}. We still go ahead and create a simple LOM with a handful of parameters to tune due to the limited experimental data. Furthermore, $\tilde \alpha$ equals twice the cold-flow combustor decay rate. The decay rate is a commonly measured parameter used in the acoustic characterization of combustors \citep{nair14,mariappan15}.   

The steady-state temperature, $\tilde T_s(x)$, sharply rises across the flame. Its value is measured by the two thermocouples (shown as black boxes in figure \ref{fig:geometry_PINN_setup}$a$) placed at the entrance and downstream of the flame. We assume $\tilde T_s$ to have a jump at the flame location. It is further assumed to be constant on either side of the flame, although the combustor walls are non-adiabatic. $\tilde q(\tilde t)$ is the total (spatially integrated) unsteady heat release rate due to the vortex-shedding process affecting the flame. It is assumed to be distributed spatially in a Gaussian manner $e^{-((\tilde x - \tilde x_c)/\tilde b)^2}/\tilde b\sqrt{\pi}$, where $\tilde b$ equals $\sqrt{2}$ times the standard deviation. The quantity $\tilde b$ is a measure of the flame length. Our measured flame lengths (along the $\tilde x$ direction) are around $10$ mm \citep{singh2021experimental}. We, therefore, choose $\tilde b=5$ mm. 

Before we move further, non-dimensionalization is performed with respect to the following acoustic scales: $x=\tilde x/\tilde l,~t=\tilde t/(\tilde l/\tilde c_s),~u=\tilde u/\tilde u_s,~p=\tilde p/\gamma_s M_s \tilde p_s$. The subscript $s$ indicates the steady-state values measured at the exit of the premixed burner (marked as green circles in figure \ref{fig:geometry_PINN_setup}$a$). $\tilde l$ is the axial length of the combustor. In the present case, $\tilde q$ is optically measured from the CH\textsuperscript{*} radical chemiluminescence emitted by the flame. The fluctuation in the radical emission is proportional to the unsteady heat release rate for lean flames \citep{nori09}. The data used in the present study is obtained in the combustor, which experienced lean flames during its entire operational range. A photomultiplier (PMT, blue shaded equipment in figure \ref{fig:geometry_PINN_setup}$a$) device receives the total emissions $(\tilde q_{pmt})$from the flame: $\tilde q(\tilde t)=\tilde \beta \tilde q_{pmt}(\tilde t)$, where $\tilde \beta$ is another parameter. The final non-dimensionalized equations appear as
\begin{align}
E_1&=\frac{\partial u}{\partial t}+T_s\frac{\partial p}{\partial x}=0\label{eq:final_damped_wave_eq_temp_jump_mom},\\
       E_2&=\frac{\partial p}{\partial t}+\frac{\partial u}{\partial x}+\alpha p-\beta q_{pmt} \frac{e^{-\left(\left(x - x_c\right)/b\right)^2}}{b\sqrt{\pi}}=0\label{eq:final_damped_wave_eq_temp_jump_en}.
\end{align}
where $\beta=(\gamma_s-1)\tilde \beta/\gamma_s \tilde u_s \tilde p_s \tilde A$. The quantity $u_s$ is measured in the experiments, while $p_s$ is assumed to be the atmospheric pressure, as the combustor is open at both ends. Essentially, we aim to estimate the scaled parameter $\beta$, where $\beta$ represents the driving from the flame to the acoustic field. 
 
$T_s(x)$ in \eqref{eq:final_damped_wave_eq_temp_jump_en} is the steady state temperature along the length ($x$) of the combustor. As said before, the flame produces a sharp rise in $T_s$ and is modeled as a jump. Since the non-dimensionalization is performed with respect to the upstream steady-state flow variables, $T_s$ takes the following form: $T_s(x)=1+(T_r-1)H(x-x_c)$, where $T_r$ is the ratio of steady-state temperatures across the flame, and $H(x)$ is the Heaviside function to provide the jump. The discontinuity in $T_s$ does not result in the discontinuous $p,~u$ \citep{lieuwen12}. Therefore, there is no threat to the PINN's performance when the solution is continuous. $T_r$ almost remains constant in the experiments and is found to be around 2.6 (refer figure 6($c$) of \citealt{singh2021experimental}). The same value is used in the present study. The non-dimensionalized $b$ equals 0.005.

\subsection{Van der Pol oscillator model for flame dynamics}
An equation for $q_{pmt}(t)$ is required to complete the equations set \eqref{eq:final_damped_wave_eq_temp_jump_mom}-\eqref{eq:final_damped_wave_eq_temp_jump_en}. Vortex shedding affects $\tilde q$, and it occurs due to the absolute wake instability behind the bluff body \citep{huerre90}. Therefore, the formed von Karman shedding process is a self-sustained nonlinear oscillator. The Van der Pol oscillator is the simplest model to represent such an oscillator. In fact, the equation of the van der Pol oscillator can be obtained from the Navier-Stokes equation (near the first Hopf point) for vortex shedding behind the cylinder \citep{duvsek1994numerical}, indicating its physics-based nature. It is a versatile model for vortex-induced structural oscillations \citep{facchinetti2004coupling,ogink2010wake}. Modeling the flame dynamics by the van der Pol oscillator caused by a hydrodynamic instability (similar to vortex shedding)  reproduces the qualitative dynamics of thermoacoustic oscillations observed in the experiments \citep{li13a}. Therefore, the van der Pol oscillator for $q_{pmt}$ is simple yet close to a physics-based flame dynamics model. It is written in the following form:
 \begin{equation}\label{eq:van_der_pol_osc}
     E_3=\frac{d^2q_{pmt}}{dt^2}-\mu (1-q_{pmt}^2) \frac{dq_{pmt}}{dt}+\omega_v^2q_{pmt}-\gamma u(x_r,t)=0,
 \end{equation}
where $\mu$ sets the nonlinearity in the heat release rate, and $q_{pmt}$ oscillates near to the circular frequency $\omega_v$ in the absence of external perturbations. The oscillating frequency is altered by the feedback from the acoustic field through the source term $\gamma u(x_r,t)$. The velocity fluctuations at the origin of shedding affect the vortex-shedding process \citep{matveev03b}. In the present case, shedding begins from the lower side periphery ($x=x_r$) of the bluff body (figure \ref{fig:geometry_PINN_setup}$a$). Therefore, the velocity fluctuation $u(x_r,t)$ is the forcing term with an amplitude $\gamma$. $\omega_v$ is known from the experiments (refer figure 6$b$ of \citealt{singh2021experimental}). Among the three equations (\ref{eq:final_damped_wave_eq_temp_jump_mom}-\ref{eq:van_der_pol_osc}), there are four parameters: $\alpha,~\beta,~\mu$ and $\gamma$. 
 
In the present study, we aim to identify the value of the above four parameters ($\alpha,~\beta,~\mu$, and $\gamma$), along with the entire $p,~u$ (hidden variables) and $q_{pmt}$ fields. This is accomplished by using the measured total heat release of the flame ($q_{pmt,data}$) and acoustic pressure ($p_{data}$) time series acquired at three locations, along with the LOM \eqref{eq:final_damped_wave_eq_temp_jump_mom}-\eqref{eq:van_der_pol_osc}. 
\begin{figure}
\centerline{\includegraphics[keepaspectratio,width=1\textwidth]{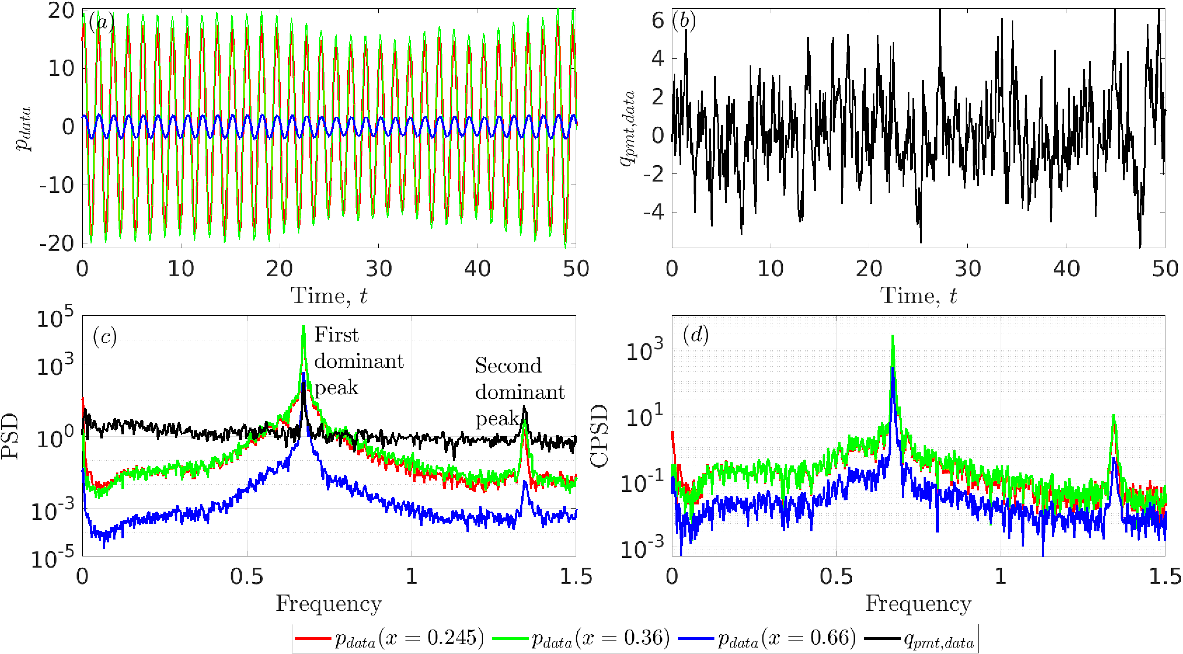}}
\caption{\textbf{Characteristics of the experimental data:} Time series of raw unsteady $(a)$ acoustic pressure and $(b)$ heat release rate. The corresponding $(c)$ power spectral (PSD) and $(d)$ cross-power spectral densities (CPSD) are in the bottom panels. Each CPSD is between pressure and heat release rate fluctuations. The airflow rate is 42 slpm, where A-lock-in occurs.}
\label{fig:time_series_turbulent_nature_42_slpm}
\end{figure}

\subsection{Characteristics of the experimental data}
Figure \ref{fig:time_series_turbulent_nature_42_slpm} shows the raw time series (top panels), along with their corresponding power spectral densities (PSD) and cross-power spectral densities (CPSD) between the recorded acoustic pressure and heat release rate fluctuations. The airflow rate is 42 slpm (standard liters per minute), where A-lock-in occurs, leading to combustion instability. The following are the important points. 1) Even during the large amplitude oscillations (combustion instability) with a well-defined periodic pattern (figure \ref{fig:time_series_turbulent_nature_42_slpm}$a$) in the pressure time series, the corresponding amplitudes vary with time in a relatively longer time scale. 2) The unsteady heat release time series (figure \ref{fig:time_series_turbulent_nature_42_slpm}$b$) shows random fluctuations due to turbulence amid the periodic oscillations due to the thermoacoustic interaction. 

On the other hand, the physics-based LOM has the following capability. 1) The acoustic equations (\ref{eq:final_damped_wave_eq_temp_jump_mom}-\ref{eq:final_damped_wave_eq_temp_jump_en}) provide the frequency of the acoustic mode. They are linear, and the amplitude modulation observed in the pressure can be provided only through the heat release source term $q_{pmt}$. 2) $q_{pmt}$ is modeled as the nonlinear van der Pol oscillator, producing a limit cycle. Since we consider the regime close to the lock-in region, one expects a limit cycle \citep{balanov08} in the experiments. The turbulent flame modulates the limit cycle amplitude violently. We, therefore, apply PINN to small temporal chunks of the time series where the oscillation approximately resembles a limit cycle. After solving for each chunk using PINN individually, we temporally stitch all the results to obtain the entire time series (discussed in \S \ref{subsec:stitching}). After trying for several chunk durations, we realized 5 non-dimensional time would be an optimum choice, as PINN performs with the best accuracy. It also implies that the model parameters, $\alpha,~\beta,~\mu$, and $\gamma$ are quasi-constant in a given chunk.
\begin{figure}
\centerline{\includegraphics[keepaspectratio,width=1\textwidth]{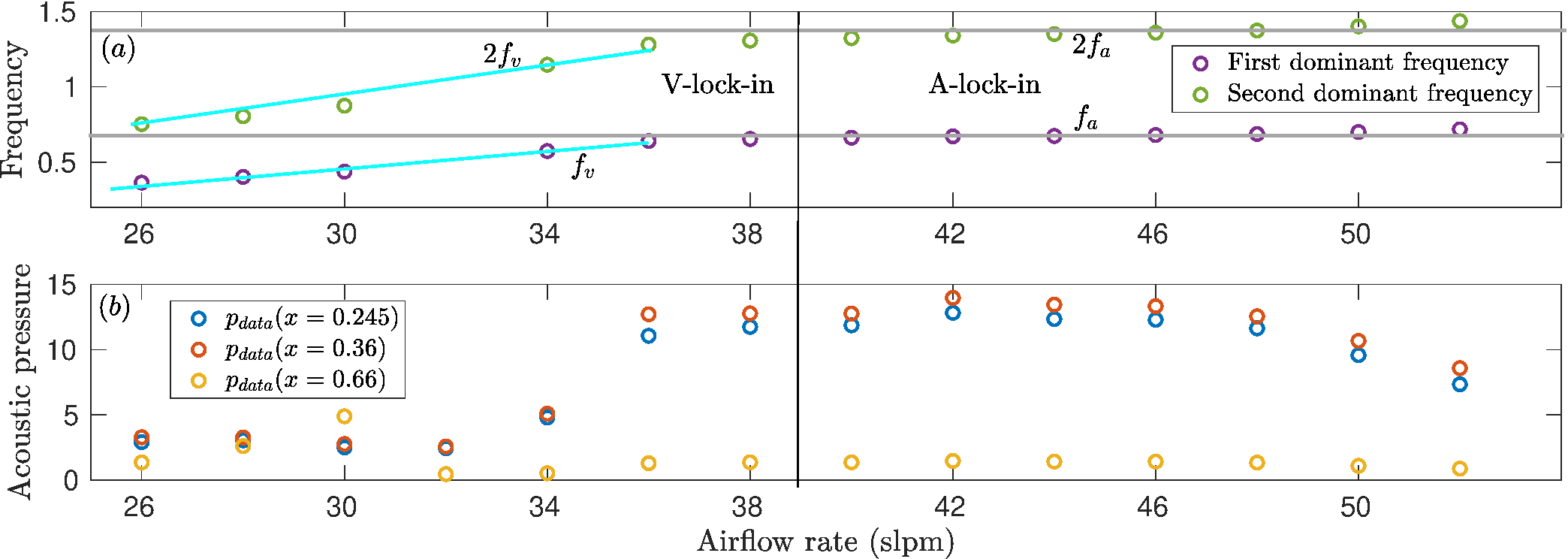}}
\caption{\textbf{V- and A-lock-in regions:} $(a)$ Variation of the first and second dominant non-dimensional frequency. $(b)$ Root mean square value of the acoustic pressure. Blue and gray lines indicate the trend lines of the vortex shedding and acoustic frequencies, respectively. The vertical black line shows an approximate demarcation between V- and A-lock-in regions.}
\label{fig:experimental_data_compilation}
\end{figure}
The lower panels of figure \ref{fig:time_series_turbulent_nature_42_slpm} show two dominant peaks close to the fundamental and first harmonic of the acoustic frequency. It is representative of the A-lock-in region. The coincidence of the frequencies at the peaks both in the PSD and CPSD reveals the periodic oscillation amid strong turbulent fluctuations of the flame. 

Such peaks are observed in the entire lock-in region. Their values are shown in the non-dimensional form as the first and second dominant frequency in figure \ref{fig:experimental_data_compilation} $(a)$. Between 26-36 slpm of airflow rate, the two dominant frequencies increase approximately in a linear trend (blue lines: $f_v$ and $f_{2v}$). There is no other peak, indicating that acoustic oscillations occur at the vortex shedding frequency: V-lock-in. At 38 slpm, vortex shedding frequency comes close to the acoustic frequency of the combustor (flat gray trend lines: $f_a$ and $f_{2a}$), indicating the onset of their interaction. Beginning from 40 slpm, the two dominant peaks closely follow the flat lines, indicating vortex shedding occurring at the acoustic frequency: A-lock-in region. Since lock-in is described based on the oscillation frequency, it is \emph{frequency} lock-in \citep{balanov08}. In this paper, lock-in indicates only frequency lock-in. The subsequent panel $(b)$ shows a strong rise in the root mean square value of the acoustic pressure fluctuations near the transition (vertical black line) between V- and A-lock-in regions. 

\subsection{Challenges for PINN}
The challenges that PINN needs to overcome for the inverse problem are multi-fold. (i) The equation set \eqref{eq:final_damped_wave_eq_temp_jump_mom}-\eqref{eq:final_damped_wave_eq_temp_jump_en} is ill-posed, as the initial condition $p(x,0)$ (except at the measurement locations), $u(x,0)$ are unknown. Boundary conditions, too, are unknown, although they are close to an acoustically open condition. (ii) Unsteady pressure measurements at a limited (three) number of spatial locations are measured without any presumption of its future use in a physics-based machine learning method. A recent paper by \cite{juniper2022generating} clearly highlights the importance of having more unsteady pressure sensors and their intelligent placements. (iii) The turbulent flow is not captured in the LOM. Finally, we note that an industrial combustor would have the above challenges, too. 

\section{PINN for the inverse problem of thermoacoustic interaction}
\label{sec:PINN_arch}
\noindent In this section, first, we briefly review the basic approach of PINN for inverse problems. Then, we formulate PINN for our specific problem of thermoacoustic interaction discussed in section \ref{sec:lom}.

\subsection{PINN for inverse problem}
PINN considers two loss terms, namely data and physics loss \citep{raissi2019physics}. The former is a function of the difference in the measured and predicted data. The latter is the residue of the physical equation (ODEs/PDEs) that needs to be zero in the domain. The derivatives involved in residue calculation are obtained using automatic differentiation \citep{Baydin_2018}. PINN is suitable for both forward and inverse problems.

In figure \ref{fig:PINN}, we show a schematic diagram of PINN consisting of two parts. The first part, shown on the left-hand side, is a deep neural network (DNN), which takes $x,~t$ as input and approximates an output $\bar{v}$. DNN is indicated as $\mathcal{N}(\bm{\theta};x,t)$, where $\bm{\theta}$ represents weights ($\bm{W}$) and biases ($\bm{b}$) of the network (the parameters of the neural network.) The equation for the DNN with 2 input neurons (for $x,~t$) having $n$ hidden layers terminating to an output neuron, producing $\bar{v}$ can be written as 
\begin{subequations}
    \begin{align}
        \bm{v}_0 & = \{x, t\} \hspace{6cm} && \text{Input}\\
        \bm{v}_j & = \sigma\left(\bm{W}_j\bm{v}_{j-1} + \bm{b}_j\right) \hspace{1cm} j\;\forall \;1\le j\le n - 1 \hspace{0.5cm} && \text{Hidden layers}\\
        \bar v = v_n & = \bm{W}_n\bm{v}_{n-1} + \bm{b}_{n-1} && \text{Output layer}
    \end{align}
\end{subequations}
where $\bm{W}_j$ and $\bm{b}_j$ are the weights and biases of the $j$\textsuperscript{th} layer. All the $\bm{W}_j$s and $\bm{b}_j$s form the network parameters $\bm{\theta}=\{\bm{W}, \bm{b}\}$. $\sigma(.)$ is the activation function for the hidden layers. In the present study, we employ the $\tanh(.)$ activation function. 
\begin{center}
\includegraphics[width=0.95\textwidth]{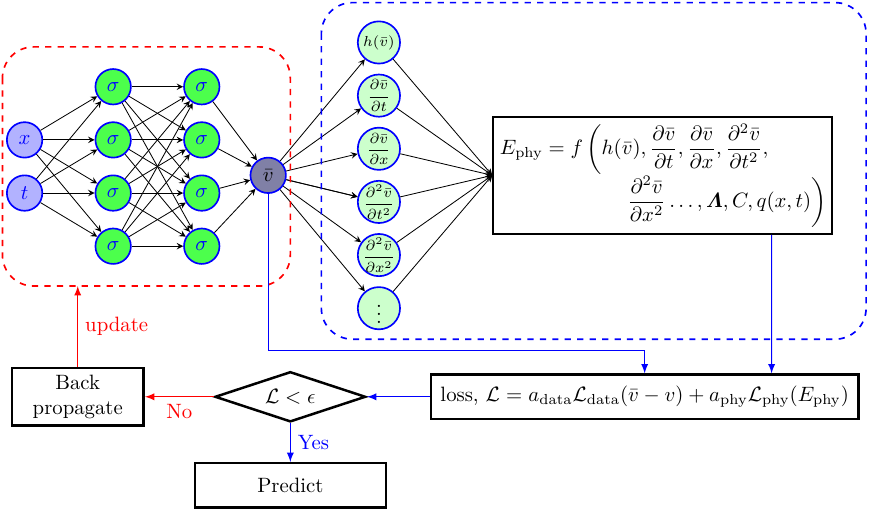}
\captionof{figure}{\textbf{Schematic of PINN for inverse problems:} A schematic diagram of PINN showing a DNN on the left side with two input ($x,~t$) neurons, providing an output $\bar{v}$. The right side shows the physics of the problem where the residue of the PDE is calculated. The form of the PDE is $f\left(h(\bar{v}), \frac{\partial\bar{v}}{\partial t}, \frac{\partial\bar{v}}{\partial x}, \frac{\partial^2\bar{v}}{\partial t^2}, \frac{\partial^2\bar{v}}{\partial x^2}, \dots, \bm{\varLambda}, C, q(x, t) \right)=0$; $h$ is a function of the state variable ($\bar v$) without differentiation. $\bm{\varLambda}$ and $\bm{C}$ are the unknown parameters and the known coefficients/constants, respectively, of the equation; $q(x,t)$ is the driving force, if any, involved in the PDE. The total loss $(\mathcal{L})$ consists of the physics and data loss; the latter comprises the loss associated with the boundary, initial conditions, and measured data. $\bm{\varLambda}$ are trained, along with $\bm{\theta}$ in the optimization process.}
\label{fig:PINN}
\end{center}

The right side of figure \ref{fig:PINN} shows the calculation of the residue ($E_{\text{phy}}$) of the ordinary/partial differential equation (ODE/PDE). Automatic differentiation \citep{Baydin_2018} is employed to calculate the derivatives. The unknown parameters of the PDE are grouped in $\bm{\varLambda}$. The residue is used to calculate the physics loss $\mathcal{L}_{\text{phy}}(E_\text{phy},\bm{\varLambda})$. 

The data loss $\mathcal{L}_{\text{data}}(v_n-v_{data})$ comprises the loss associated with the initial, boundary conditions and experimental data. Additionally, we use self-adaptive weights \citep{mcclenny2020self} in both the physics ($\bm{\lambda}_{phy}$) and data ($\bm{\lambda}_{data,b}$) loss terms for each of the residual and data points. These weights aid in faster convergence of the optimization problem described soon. The total loss is the weighted sum of the above two loss functions.
\begin{equation}
    \mathcal{L}(\bm{\theta}, \bm{\varLambda}, \bm{\lambda}_{phy}, \bm{\lambda}_{data}) = a_{phy}\mathcal{L}_{\text{phy}}(\bm{\theta}, \bm{\varLambda}, \bm{\lambda}_{phy}) + a_{data}\mathcal{L}_{\text{data}}(\bm{\theta}, \bm{\lambda}_{data}),
\end{equation}
where $a_{phy}$ and $a_{data}$ are the user defined loss weights. The unknowns are the network parameters ($\bm{\theta} = \{\bm{W},\bm{b}\}$), ODE/PDE parameters ($\bm{\varLambda}$) and self-adaptive weights ($\bm{\lambda}= (\bm{\lambda}_{phy},~\bm{\lambda}_{data})$). They are determined by posing a min-max optimization problem. The loss function $\mathcal{L}$ is minimized with respect to $\bm{\theta},~\bm{\varLambda}$, while it is maximized with respect to $\bm{\lambda}$. It is written as
\begin{equation}\label{eq:min_max_prob}
    \underset{\bm{\theta}, \bm{\varLambda}}{\min}\;\; \underset{\bm{\lambda}}{\max} \;\;\mathcal{L}(\bm{\theta}, \bm{\varLambda}, \bm{\lambda}).
\end{equation}
Considering a gradient-based optimization, the optimization process may be written as
\begin{equation}\label{eq:opt_proc_1}
    \bm{\theta}, \bm{\varLambda} = \bm{\theta}, \bm{\varLambda} - lr_{\bm{\theta}, \bm{\varLambda}} \nabla_{\bm{\theta}, \bm{\varLambda}}\mathcal{L}(\bm{\theta}, \bm{\varLambda}, \bm{\lambda}),
\end{equation}
\begin{equation}\label{eq:opt_proc_2}
    \bm{\lambda} = \bm{\lambda} + lr_{\bm{\lambda}} \nabla_{\bm{\lambda}}\mathcal{L}(\bm{\theta}, \bm{\varLambda}, \bm{\lambda}),
\end{equation}
where $lr_{\bm{\theta}, \bm{\varLambda}}$ is the learning rate associated with $\bm{\theta}$ and $\bm{\varLambda}$, while $lr_{\bm{\lambda}}$ is the learning rate associated with $\bm{\lambda}$. In the present study, we use multiple Adam optimizer \citep{kingma2017adam} in Tensorflow-2 with 32-bit precision to obtain the optimal parameters. In the case of multiple physical equations (ODEs/PDEs) or/and multiple known (data) quantities, the physics and data losses are the weighted sum of individual physics and data losses. Furthermore, in the case of an unknown field instead of scalar quantity, it can be represented using a neural network.

\subsection{PINN formulation for the thermoacoustic interaction}
\label{subsec:pinn_phil}
For the current problem, we approximate the state variables $(p,~u,~q_{pmt})$ using three neural networks as shown in Table \ref{tab:network size}. The total loss function $\mathcal{L}$ is defined as the weighted sum (with known weights) of individual loss functions, given by \eqref{eq:loss_fn_gen}. 
\begin{center}
\begin{tabular}{C{1cm}C{3cm}C{1cm}C{3cm}C{2cm}} \cline{1-5}
\# & Neural network & Input  & Network size & Output \\ \cline{1-5}
1 & $\mathcal{N}_1(\bm{\theta}_1;x,t)$ & $x$, $t$ & $2-500\times3-1$ & $p(x,t)$  \\ \cline{1-5}
2 & $\mathcal{N}_2(\bm{\theta}_2;x,t)$ & $x$, $t$ & $2-500\times3-1$ & $u(x,t)$  \\ \cline{1-5}
3 & $\mathcal{N}_3(\bm{\theta}_3;t)$ & $t$ & $1-50\times3-1$ & $q_{pmt}(t)$  \\ \cline{1-5}
\end{tabular}%
\captionof{table}{\textbf{Network detail:} 
DNNs $\mathcal{N}_i(\bm{\theta}_i;x,t)$, $i=1,2,3$ considered for the approximation of the state variables $p(x,t)$, $u(x,t)$ and $q_{pmt}(t)$ respectively. The networks are parameterized by $\bm{\theta}$, which include both weights and biases of the neural network. The second and third columns indicate the neural network and the input variables for each network, respectively. The first two DNNs have two inputs, $x$ and $t$, while the third network has a single input, $t$. The inputs are scaled to $[-1,1]$. The fourth column indicates the network architecture. $p(x,t)$, $u(x,t)$ networks contain three hidden layers of 500 neurons each ($500\times3$). $q_{pmt}(t)$ network also has three hidden layers, containing 50 neurons in each ($50\times3$). The last column indicates the approximated state variable by each network.}\label{tab:network size}
\end{center}
\begin{equation}
\begin{split}
    \mathcal{L}(\bm{\theta}, \bm{\varLambda}, \bm{\lambda}_1,... \bm{\lambda}_9) =\: & a_{E_1}\mathcal{L}_{\text{phy}}^{(E_1)}(\bm{\theta}, \bm{\varLambda}, \bm{\lambda}_1) + a_{E_2}\mathcal{L}_{\text{phy}}^{(E_2)}(\bm{\theta}, \bm{\varLambda}, \bm{\lambda}_2) +  \\ 
    & a_{E_3} \mathcal{L}_{\text{phy}}^{(E_3)}(\bm{\theta}, \bm{\varLambda}, \bm{\lambda}_3) + 
    \\ & a_{bc_1}\mathcal{L}_{\text{bc}}^{(1)}(\bm{\theta}_1, \bm{\lambda}_4) +
    a_{bc_2}\mathcal{L}_{\text{bc}}^{(2)}(\bm{\theta}_1, \bm{\lambda}_5) +
    \\ & a_{ic_1}\mathcal{L}_{\text{ic}}^{(1)}(\bm{\theta}_1, \bm{\lambda}_6) +
    a_{ic_2}\mathcal{L}_{\text{ic}}^{(2)}(\bm{\theta}_2, \bm{\lambda}_7) + 
    \\ & a_{p_{data}}\mathcal{L}_{\text{data}}^{(p_{data})}(\bm{\theta}_1, \bm{\lambda}_8) + a_{q_{data}}\mathcal{L}_{\text{data}}^{(q_{data})}(\bm{\theta}_3, \bm{\lambda}_9) + 
    \\ & a_{u_{nn}}\mathcal{L}_{\text{nn}}^{(u_{nn})}(\bm{\theta}_2)
\end{split}\label{eq:loss_fn_gen}
\end{equation}
where $\bm{\theta} = \{ \bm{\theta}_1, \bm{\theta}_2, \bm{\theta}_3 \}$. They represent the weights and biases for the networks of $p,~u,~q_{pmt}$. The subscript ``phy" corresponds to the physics loss calculated from the residue of the equations \eqref{eq:final_damped_wave_eq_temp_jump_mom}-\eqref{eq:van_der_pol_osc}. Subscripts ``bc'' and ``ic'' are associated with the acoustic boundary and initial conditions. The subscript ``data'' represents data obtained from either simulations or experiments. The last loss term $\left(\mathcal{L}_{\text{nn}}^{(u_{nn})}\right)$ in \eqref{eq:loss_fn_gen} is a special loss since there is no measured data available for acoustic velocity. Expressions for the individual loss terms in \eqref{eq:loss_fn_gen} are given towards the end of the section.

Despite significant advancements, training a quantitatively accurate and robust neural network is still daunting. Therefore, we perform different cases in the order of increasing complexity. Simulated data from the numerical simulations are used in the first four cases before moving to the usage of experimental data. It is important that we consider the cases with simulated data before moving to the experimental data. This way, we can verify the PINN model and check for prediction accuracy, as the simulated data are the exact solution of the LOM. The performance of the PINN can be evaluated by comparing its solution with the simulated one in the entire $x,~t$ domain. The following five cases are considered in order of complexity. The fourth case is an imitation of the experimental case with simulated data.
\vspace{0.5cm}
\begin{enumerate}[label=\textbf{Case \arabic* : }, leftmargin=*, itemindent=-0pt]
\setlength{\itemsep}{6pt}
    \item  Acoustic wave equations (\ref{eq:final_damped_wave_eq_temp_jump_mom}-\ref{eq:final_damped_wave_eq_temp_jump_en}) in the presence of a passive flame ($\beta=0$, causing only $T_s$ rise) is solved for various known $\alpha$, initial $(p(x,0),~u(x,0))$ and boundary $(p(0,t),~p(1,t))$ conditions using PINN. The entire $p,~u$ field forms the output. They are compared with the analytical solution. This is a well-posed-forward problem. This study is the first step toward solving the thermoacoustic problem. The numerical results are discussed in \S \ref{sec:sol_acoustic_eq}.
    
    \item  In the subsequent section (\S \ref{sec:synthetic_data}), data from synthetic acoustic pressure time series using the analytical solution is used as a surrogate for the experimental data. The data is provided at the exact locations and sampling frequency as in the experiments. Boundary conditions are provided, while there are no initial conditions. A comparison of the PINN solution with the analytical result ensures that PINN can solve an ill-posed problem with the data. In this case, $\beta=0$ and $\alpha$ are also known. 
    
    \item In this case (discussed in \S \ref{sec:heaT_source_inclusion}), the heat source $q_{pmt}$ (active flame) is also included in the simulation and PINN. Similar to the previous case, synthetic data for acoustic pressure ($p$) and heat release rate ($q_{pmt}$) fluctuations are produced by a numerical solution of (\ref{eq:final_damped_wave_eq_temp_jump_mom}-\ref{eq:van_der_pol_osc}). All the four parameters $\alpha,...\gamma$ are known. Other information about the boundary and initial conditions remains the same as that of case 2. The PINN solution for $p,~u$ and $q_{pmt}$ are compared with the numerical solution. 
    
    \item In this case (discussed in \S \ref{sec:inv_prob_syn}), the four LOM parameters ($\bm{\varLambda}=\{\alpha$, $\beta$, $\mu$, $\gamma\}$) are considered as unknowns. All other inputs remain the same as in case 3. The known quantities are the boundary conditions, acoustic pressure ($p$) measured at three spatial locations, and the heat release rate ($q_{pmt}$). This case imitates the experimental data. A successful performance of PINN ensures its applicability to the experimental data. 
    
    \item Case 4 is repeated using the experimental data as the final case (\S \ref{sec:inv_prob_exp}). The temporal stitching operation is performed to obtain the entire $p,~u,~q_{pmt}$ fields, along with the LOM parameters. Furthermore, the test is repeated for various airflow rates to obtain physical insights. 
     \vspace{0.5cm}
\end{enumerate}
In cases 1-4, where simulated data is used, the PINN computed solution (subscripted $nn$: abbreviated for neural network) and the predicted parameters (subscripted $nn$) are compared with the simulated solution (subscripted $si$: abbreviated for simulation) in the entire field $x,~t$ and its parameters (subscripted $si$). We interchangeably use the words ``simulated'' and ``synthetic'' to convey the same meaning.

With the terminologies defined, we write the individual loss terms of equation \eqref{eq:loss_fn_gen}. The physics loss terms corresponds to equations \eqref{eq:final_damped_wave_eq_temp_jump_en}-\eqref{eq:van_der_pol_osc} are defined as
\begin{equation}
    \mathcal{L}_{\text{phy}}^{(E_g)} = \dfrac{1}{M_g}\sum_{m=1}^{M_g} \mathcal{F}(\lambda_g^{(m)})E^2_g(\bm{\theta};x, t)_m,\quad g=1,2,3
\end{equation}
where $E_g$, $g=1,2,3$ are the residue at each discretized point (sub/superscripts $m$) in $x,t$ domain, for equations \eqref{eq:final_damped_wave_eq_temp_jump_mom}-\eqref{eq:van_der_pol_osc}, respectively. $M_g$ is the total number of spatiotemporal residual points; $\lambda^{(m)}_g$ are the self-adaptive weights at each residual point; $\mathcal{F}$ is a mask function considered on the self-adaptive weights $\lambda^{(m)}_g$, so that the weights are always positive. In the present study, we consider the soft-plus mask function for all the self-adaptive weights. The boundary losses in equation \eqref{eq:loss_fn_gen} are defined as 
\begin{equation}
    \mathcal{L}_{\text{bc}}^{(g)} = \dfrac{1}{N_t}\sum_{n=1}^{N_t} \mathcal{F}(\lambda_{g+3}^{(n)})[p_{nn}(x_g,t)_n - p_{si}(x_g,t)_n]^2,\quad g=1,2
\end{equation}
where $g=1,2$ indicates the acoustic boundary condition at the entrance and exit of the combustor. Since both ends are open to the atmosphere, $p_{si}(x_g,t)_n=0$ at $x_1=0$ and $x_2=1$. $N_t$ is the total temporal discretized (sub/superscripts $n$) points. $\lambda_{g+3}^{(n)}$ are the self-adaptive weights at each boundary points. The initial condition in the loss equations \eqref{eq:loss_fn_gen} are defined as,
\begin{equation}
\begin{split}
   \mathcal{L}_{\text{ic}}^{(1)} &= \dfrac{1}{N_x}\sum_{n=1}^{N_x} \mathcal{F}(\lambda_6^{(n)})[p_{nn}(x,t=0)_n - p_{si}(x,t=0)_n]^2\\
      \mathcal{L}_{\text{ic}}^{(2)} &= \dfrac{1}{N_x}\sum_{n=1}^{N_x} \mathcal{F}(\lambda_7^{(n)})[u_{nn}(x,t=0)_n - u_{si}(x,t=0)_n]^2
\end{split}
\end{equation}
The loss is associated with the initial conditions of $p,~u$. $N_x$ is the total spatial discretized points. The last set of terms, loss due to the data, is as follows.
\begin{equation}
    \mathcal{L}_{\text{data}}^{(p_{data})} = \dfrac{1}{3N_d}\sum_{n=1}^{3N_d} \mathcal{F}(\lambda_{8}^{(n)})[p_{nn}(t)_n - p_{data}(t)_n]^2
\end{equation}
where the subscript ``data'' can be either the simulated (cases 2-4) or experimental (case 5) data. Acoustic pressure is measured at three locations in the experiments. The locations are $x_1=0.245,~x_2=0.36,~x_3=0.66$. $N_d$ is the total number of discrete time instances when the data is acquired. $p_{data}$ is constructed by stacking the individual pressure time series from the three spatial locations as a single-column vector of size $3N_d\times 1$.

Similarly, the loss term due to heat release is as follows.
\begin{equation}
    \mathcal{L}_{\text{data}}^{(q_{data})} = \sum_{n=1}^{N_d} \mathcal{F}(\lambda^{(n)}_{9})[q_{pmt,nn}(t)_n - q_{pmt,data}(t)_n]^2.
\end{equation}

The acoustic velocity $u$ is not measured. In our experiments, we observed (even during limit cycle oscillations) that the time average of the acoustic pressure over at least one acoustic cycle is close to zero. It allows us to reasonably assume that time-averaged $u$ over one (or many) acoustic cycle(s) is also close to zero (no acoustic streaming). The assumption makes the PINN retain the bias of the last layer.
\begin{equation}
    \mathcal{L}_{\text{nn}}^{(u_{nn})} = \frac{|\sum_{m=1}^{M_g}u_{nn}(x,t)_{m}|}{{\rm{max}}|u_{nn}(x,t)_m|},
\end{equation}
where $|~.~|$ and ``max'' indicate the absolute and maximum operations, respectively. It is true that the considered time chunk of 5 non-dimensional time may not have an integer number of cycles. However, in the absence of measured $u$, the above approximation is the best that we could make. The loss weight $a_{u_{nn}}$ in the loss function is cautiously kept one order lower than the weights of the governing equations (Table \ref{tab:weights_case}) to ensure that this approximation does not produce undue errors. In spite of this approximation, we observed that PINN is able to capture the solution with less than 5\% (test) error in all the tests related to numerical simulation (cases 1-4).

The weights of the DNN are initialized from a random Gaussian distribution described in \cite{glorot2010understanding}. The biases are initialized with zeros. All the self-adaptive weights $\bm{\lambda}$ are initialized with unity. In case 4, where $\bm{\varLambda}$ are also unknowns, they are initialized from a Gaussian distribution with mean as the true value and standard deviation as 40\% of the true value. Ten independent runs are performed to accumulate the statistics due to the random initialization. 

After several runs in each case, we found the optimum values of various hyperparameters (learning rates scheduler and weights in the loss function $a_i$). They are tabulated in Appendix \ref{app:hyperparameters}. As discussed earlier, the optimal parameters of the problem are evaluated using multiple Adam optimizers with a total of eight hundred thousand iterations. 

The performance of PINN is evaluated by comparing its solution with the simulated solution (cases 1-4) over the entire domain $x,~t$ using test error ($E_r$). It is the average $L_2$-norm of the difference normalized with the $L_2$-norm of the simulated solution, evaluated in $\%$: 
\begin{equation}
    E_r=\dfrac{||h_{nn}-h_{si}||_2}{||h_{si}||_2}\times 100 \% .
\end{equation}
Depending on the context, $h$ stands for the acoustic pressure, velocity, or heat release rate. Test error is calculated at twice the number of residual points. The above expression is also used to evaluate the train error.

\section{Numerical results: case 1 - well-posed forward problem for the acoustic wave equation}\label{sec:sol_acoustic_eq}
We first begin with the well-posed case of solving the acoustic wave equation for the passive flame. We will compare analytical and PINN solutions.

\subsection{Analytical solution}\label{subsec:anal_eq}
For $\beta=0$, equations (\ref{eq:final_damped_wave_eq_temp_jump_mom}-\ref{eq:final_damped_wave_eq_temp_jump_en}) form an eigenvalue problem, which can be solved analytically. Since the problem is linear, we write $p=\Re\left(\hat p e^{i\omega t}\right),~u=\Re\left(\hat u e^{i\omega t}\right),~i=\sqrt{-1}$, where $\hat p(x)$ and $\hat u(x)$ form complex functions of $x$, termed as \emph{pure} acoustic modes. The term ``pure'' is used as the flame is passive (only raises the steady state temperature $T_s$). The acoustically open boundary condition at the ends, $p(x=0,t)=p(x=1,t)=0$ is used. Multiple (complex) eigenvalues $\omega$ are obtained by solving the following nonlinear algebraic equations.
\begin{align}\label{eq:disp_eq}
i\sqrt{T_r}\sin{(kx_c)}\left[e^{ikx_c/\sqrt{T_r}}+e^{ik\left(2-x_r\right)/\sqrt{T_r}}\right]&=\cos{(kx_c)}\left[e^{ikx_c/\sqrt{T_r}}-e^{ik\left(2-x_r\right)/\sqrt{T_r}}\right]\\
\omega&=\frac{i\alpha}{2}+\sqrt{k^2-\left(\frac{\alpha}{2}\right)^2}
\end{align}
where $k$ is the wave number. Real and imaginary parts $\omega$ represent the oscillating frequency ($f_{a}=\Re(\omega)/2\pi $) and the damping $(\alpha/2)$ of the acoustic waves. Modes are numbered in the order of increasing $f_a$. Expressions for $\hat{p}$ and $\hat{u}$ are written as follows.
\begin{equation}
\hat p = \begin{cases}
2A_Ii \sin(kx),  &0<x\leq x_c\\
-A_{II}\left[e^{ikx/\sqrt{T_r}}-e^{ik\left(2-x\right)/\sqrt{T_r}}\right], &x_c<x<1
\end{cases}
\end{equation}
\begin{equation}
\hat u = \begin{cases}
-2A_I\frac{k}{\omega} \cos(kx), &0<x\leq x_c\\
-A_{II}\frac{\sqrt{T_r}k}{\omega}\left[e^{ikx/\sqrt{T_r}}+e^{ik\left(2-x\right)/\sqrt{T_r}}\right], &x_c<x<1
\end{cases}
\end{equation}
where $A_{II}=2A_Ii \sin(kx_c)/\left[e^{ikx_c/\sqrt{T_r}}-e^{ik\left(2-x_c\right)/\sqrt{T_r}}\right]$. 

\subsection{PINN solution}
Equations \eqref{eq:final_damped_wave_eq_temp_jump_mom}-\eqref{eq:final_damped_wave_eq_temp_jump_en} are solved using PINN. Initial conditions for $p,~u$ are generated by assuming $2A_Ii=1$, as they form an eigenvalue problem. $\alpha$ is the only (known) parameter which is varied. A total of 101 and 105 equally spaced discretized points in the $x$ and $t$ directions, respectively, are used for residual points in the calculation of physics loss. It is a well-posed-forward problem. Therefore, we consider the loss weights $a_{E_1}=a_{E_2}=a_{bc_1}=a_{bc_2}=a_{ic_1}=a_{ic_2}$ equal (set to unity). Since only acoustic equations are solved $a_{E_3}=0$, furthermore, no data are involved, $a_{p_{data}}=a_{q_{data}}=a_{u_{NN}}=0$ in the loss function $\mathcal{L}$ \eqref{eq:loss_fn_gen} 

\begin{figure}
\includegraphics[keepaspectratio,width=1\textwidth]{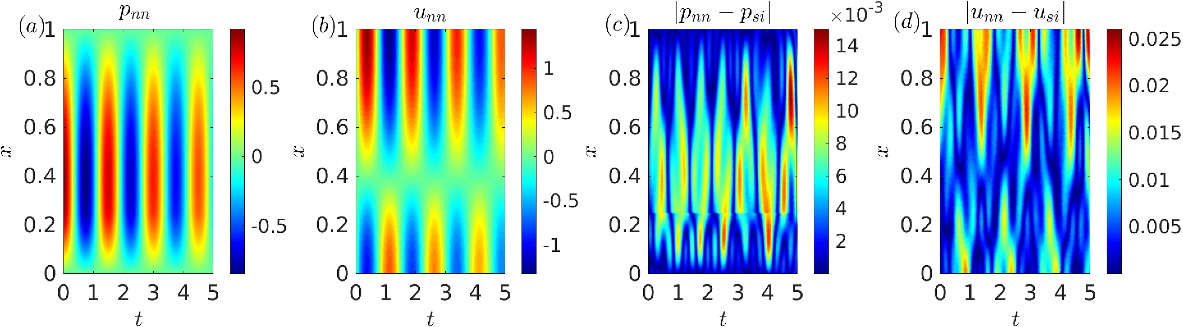}\\
\includegraphics[keepaspectratio,width=1\textwidth]{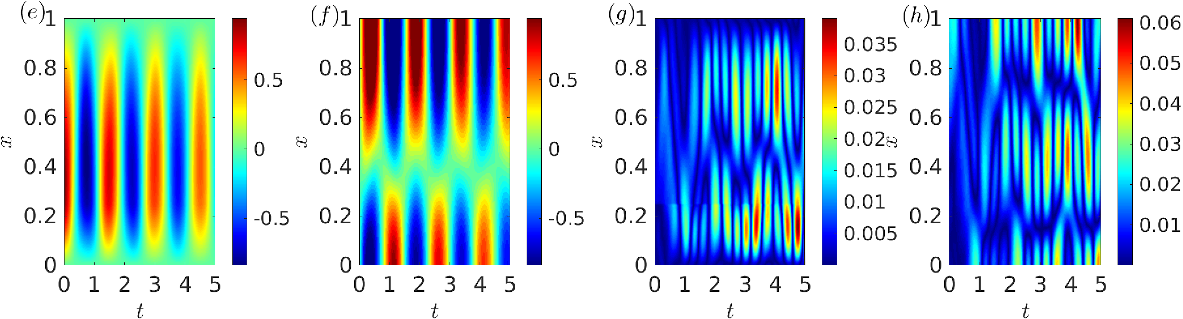}
\caption{\textbf{Case 1 - study of PINN and simulated results:} Figure shows the comparison of the PINN solution $(p_{nn},~u_{nn})$ with the analytical $(p_{si},~u_{si})$ solution. The top row contains the results when only the fundamental mode is excited. The bottom one corresponds to the excitation of both the fundamental and first harmonic acoustic modes. The first two columns indicate the solution from PINN. The last two columns show the absolute difference between the PINN and the analytical solutions.}
\label{fig:wave_eq_2_68Tr_alpha_0_2}
\end{figure}

Figure \ref{fig:wave_eq_2_68Tr_alpha_0_2} shows the comparison of the PINN solution $(p_{nn},~u_{nn})$ with the analytical $(p_{si},~u_{si})$ solution. As a first step, only the fundamental acoustic mode (top panel), which is found to be the dominant mode in the experiments, is excited. Mode shapes of acoustic pressure $(p_{nn})$ and velocity $(u_{nn})$ (first two columns) are captured well. Acoustic damping in their amplitudes is also seen. The last two columns indicate the point-wise absolute of their differences with the analytical solutions $(p_{si},~u_{si})$. The point-wise differences are low. For the realization in figure \ref{fig:wave_eq_2_68Tr_alpha_0_2}, the test error $E_r$ amounts to 1.47 \% and 1.78 \% for acoustic pressure and velocity, respectively. It indicates a close match with the analytical solution. In this case, we varied the number of layers and neurons per layer for the neural network (results are not shown). This test allowed us to land in the optimum neural network architecture for $p,~u$ (first two rows of Table \ref{tab:network size}). The mean and standard deviations over ten independent realizations are shown in the first two rows of Table \ref{tab:PINN_drror}. 

In our experiments (\S \ref{sec:inv_prob_exp}), the amplitude of the first harmonic of $p$ is up to 0.06 times its fundamental acoustic mode. Therefore, we employ PINN to obtain forward solutions (bottom panels), considering the Fourier amplitude of the first harmonic to be 0.07 of the fundamental acoustic mode. The inclusion of the harmonic raises the error. In all the cases, the test error is less than 5\%, and the standard deviation is one order less than the mean (third and fourth rows of Table \ref{tab:PINN_drror}). The same is observed for multiple values of $\alpha$ (first four rows of Table \ref{tab:PINN_drror}). This indicates the robustness of the chosen neural network architecture and its hyperparameters.      

\section{Numerical results: case 2 - ill-posed acoustic wave equation using synthetic data}\label{sec:synthetic_data}
In this case, we solve the acoustic wave equation with the aid of synthetic data. The initial condition associated with $u$ is unavailable. The same is true for $p$, except for the three measurement locations. It forms an ill-posed problem. The ill-posedness is leveraged by the synthetic data generated from the analytical solution (\S \ref{sec:sol_acoustic_eq}). In this way, the performance of PINN in recovering the $p,~u$ field is evaluated. The (non-dimensional) temporal sampling rate equals 11.9. The value is chosen to be the same as in the experiments. The oscillating frequencies of the fundamental and first harmonic acoustic modes are 0.66 and 1.46, indicating that they are well-time-resolved with the sampling rate. 
 
The loss weights (in \eqref{eq:loss_fn_gen}) associated with the acoustic equations and boundary conditions are set as in case 1 to unity (second row of Table \ref{tab:weights_case}). $a_{E_3}$ still remains zero. Since the initial conditions are unknown, their weights are zero ($a_{ic_1}=a_{ic_2}=0$) too. The synthetic acoustic pressure time series $(p_{data})$ forms the data loss of the PINN. We also choose the weight $a_{p_{data}}$ to unity. The synthetic data are obtained at the same location as in the experiments (\S \ref{sec:combustor_description}) 

Since $\beta=0$, we use \eqref{eq:final_damped_wave_eq_temp_jump_en} to obtain another constraint ${\partial p_{data}}/{\partial t}+{\partial u_{nn}}/{\partial x}+\alpha p_{data}=0$ in the loss function. This provides additional information, thereby improving the test error of the PINN. The corresponding weight $a_{q_{data}}$ is also set to unity. As said before in \S \ref{subsec:pinn_phil}, $a_{u_{nn}}=0.1$ is used to leverage the absence of acoustic velocity measurement.  
 
\begin{figure}
\includegraphics[keepaspectratio,width=1\textwidth]{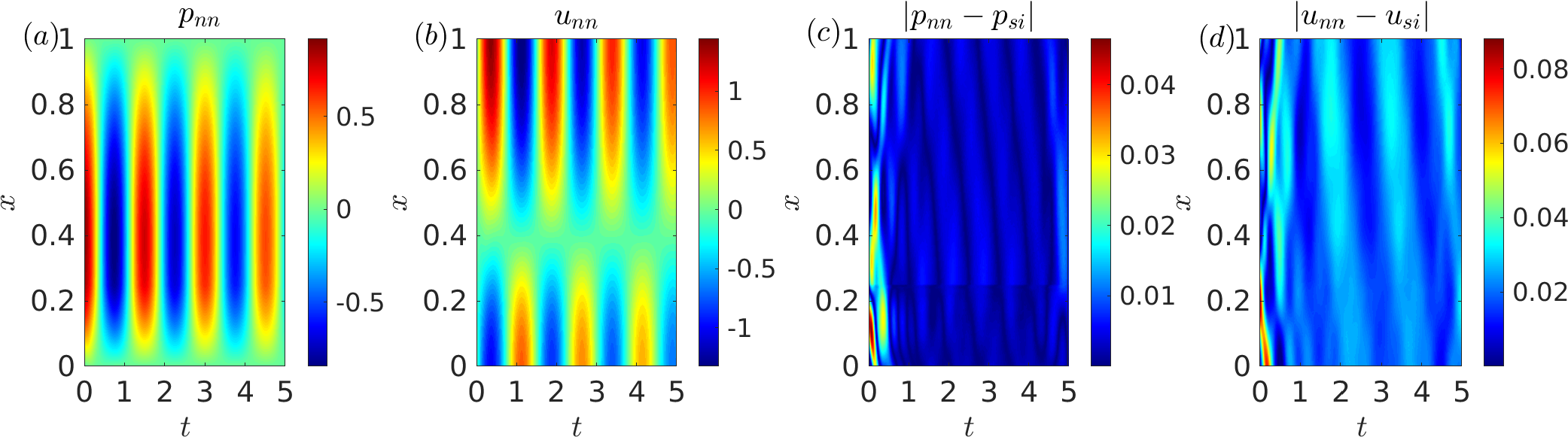}\\
\caption{\textbf{Case 2 - study of PINN and simulated results:} Figure shows the comparison of the PINN solution $(p_{nn},~u_{nn})$ with the analytical $(p_{si},~u_{si})$ solution. The first two panels indicate the solution from PINN. The last two panels show the absolute difference between the PINN and the analytical solutions.}
\label{fig:wave_eq_2_68Tr_alpha_0_2_synthetic_data}
\end{figure}

 The results in figure \ref{fig:wave_eq_2_68Tr_alpha_0_2_synthetic_data} show a good match with the analytical solution again. The test error in $u$ is almost double that for $p$ (last two rows of Table \ref{tab:PINN_drror}). It is not surprising as information only on $p$ is fed. In any case, both the test errors are less than 5\%. Measuring $u$ directly in a combustor is challenging, especially the practical ones operating at high pressures. The PINN framework allows for obtaining it with reasonable accuracy. Furthermore, although the input is only $p$ time-series from a few spatial locations, the entire $u,~p$ field is obtained, which adds to the applicability of PINN to the thermoacoustic system.     
 
\section{Numerical results: case 3 - ill-posed problem of thermoacoustic interaction using synthetic data}\label{sec:heaT_source_inclusion}
The next step is to apply PINN on the three coupled equations (\ref{eq:final_damped_wave_eq_temp_jump_mom}-\ref{eq:van_der_pol_osc}) representing the thermoacoustic interactions. Synthetic data for $p_{data}$ at three locations and $q_{data}$ are generated by solving them numerically. All the four LOM parameters $\alpha,...\gamma$ are known. An important parameter during lock-in is the frequency ratio ($f$), which is the ratio of the two frequencies of the oscillators (acoustic field and vortex shedding) in their uncoupled state: $\beta=\gamma=0$. The acoustic frequency $f_a$ is obtained from the analytical solution (\S \ref{sec:sol_acoustic_eq}). The frequency ratio $(f)$ is defined as $f=f_a/(\omega_v/2\pi)$. For cases 3-4, we assume $f$, thereby fixing the value of $\omega_v$. $f$ in case 5, is obtained from the experimentally measured frequency (discussed in \S \ref{sec:inv_prob_exp}).

Since the wave equation is involved, we prefer solving them using Chebyshev polynomials to represent the unknowns $p,~u$ and $q_{pmt}$. The package Chebfun in Matlab is used for this purpose \citep{driscoll2014chebfun}. Initial transients of the numerical solution are ignored. Unlike the last two sections, the weights $a_{p_{data}}$ and $a_{q_{data}}$ of $\mathcal{L}$ are increased to 10. The rest of the weights are kept the same as in the previous section (\S \ref{sec:synthetic_data}). It penalizes the (synthetic) acoustic pressure and heat release rate data more than the governing equations and the boundary conditions. Higher penalization on the data allows a faster convergence of the $p_{nn},~q_{pmt,nn}$ at the locations of the (synthetic) data. This convergence guides the entire field $u,~p$ to evolve to satisfy the governing equations. We observe this trend in all the remaining cases and is discussed in \S \ref{sec:inv_prob_syn}.

\begin{figure}
\includegraphics[keepaspectratio,width=1\textwidth]{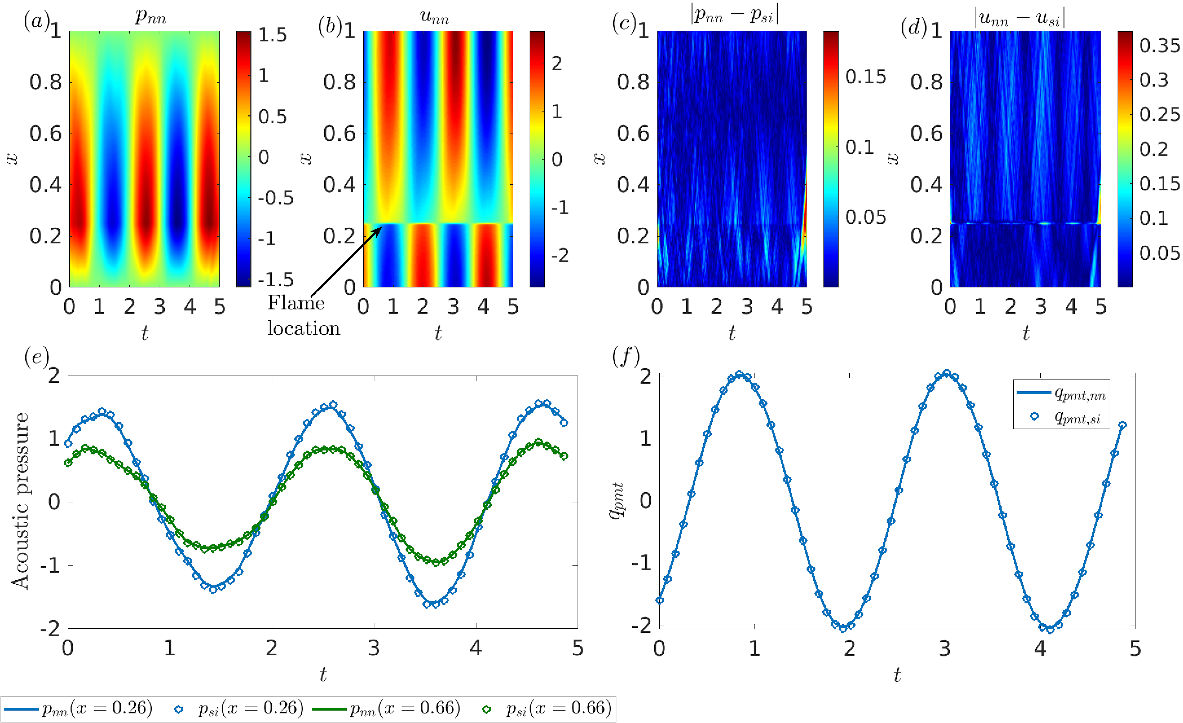}
\caption{\textbf{Case 3 - Study of PINN and simulated results:} Figure shows the comparison of the PINN solution $(p_{nn},~u_{nn})$ with the numerical $(p_{si},~u_{si})$ solution for the coupled equations (\ref{eq:final_damped_wave_eq_temp_jump_mom}-\ref{eq:van_der_pol_osc}), representing the thermoacoustic interaction. On the first row, the first two panels ($a,~b$) indicate the solution from PINN. The absolute difference between the PINN and the numerical solutions is shown in the next two panels $(c,~d)$. The bottom row panels show the evolution of the PINN solution (continuous curve) at two locations of acoustic pressure and the total heat release rate measured by the PMT. Open circles indicate the synthetic input data from the numerical solution. The parameters correspond to the first row of Table \ref{tab:thermoacoustic_problem}. The frequency ratio $f$ equals 1.1.}
\label{fig:coupled_forward_problem_param_1}
\end{figure}

Figure \ref{fig:coupled_forward_problem_param_1} shows the performance of PINN in comparison to the numerical solution. The first two panels of the top row illustrate the PINN solution of $p$ and $u$ for a realization. An important feature is the sharp change in $u$ (panel $b$) across the active flame, which is captured perfectly. The next two panels $(c,~d)$ show the point-wise difference between the PINN and numerical solutions. Most of the regions show low values. The bottom row panels $(e,~f)$ compare the time evolutions of the input synthetic data (open circles) with the PINN (continuous curve) solution. A good match in the acoustic pressure (panel $e$) and heat release (panel $f$) is observed too. The comparison for the third input acoustic pressure time series is omitted for figure clarity. 

Table \ref{tab:thermoacoustic_problem} contains the train and test errors for two parameter combinations (differing in $\gamma$). All the train errors are less than 5\%. Lower test errors are observed for $p$ and $q_{pmt}$, relative to $u$. Although the (mean) test error in $u$ is slightly above 5\%, it is reduced by grid-clustering around the flame. Clustering is implemented in the subsequent cases. 

\section{Numerical results: case 4 - parameter identification and solution to the thermoacoustic problem using synthetic data}\label{sec:inv_prob_syn}
As the final step using the synthetic data, we identify the four parameters ($\alpha,...\gamma$) and the entire $p,~u,~q_{pmt}$ field. Among the parameters, $\alpha,\beta,\mu$ are non-negative. A soft-plus mask function is applied to ensure the same. 

Since $q_{pmt}$ is one of the data that is driving the solution, we preferred spatial grid clustering around the flame location, keeping the total number of the residual points as before. A total of 21 points are clustered near the flame zone, between $x_c\pm 2b$. The remaining 80 points are distributed equally outside of the flame zone. The clustering yielded better accuracy in predicting the parameters and the field variables for the same computational cost. Furthermore, loss weights for the data ($a_{p_{data}},~a_{q_{data}}$) and the boundary conditions ($a_{bc_1},~a_{bc_2}$) are kept ten times more than the weights of the governing equations, as the solution evolves from their information. As in the previous section, synthetic data of acoustic pressure and heat release rate time series are fed to $\mathcal{L}$. The values of the parameters $\alpha,~\beta,~\mu$ and $\gamma$ are unknown. 

\begin{figure}
\includegraphics[keepaspectratio,width=1\textwidth]{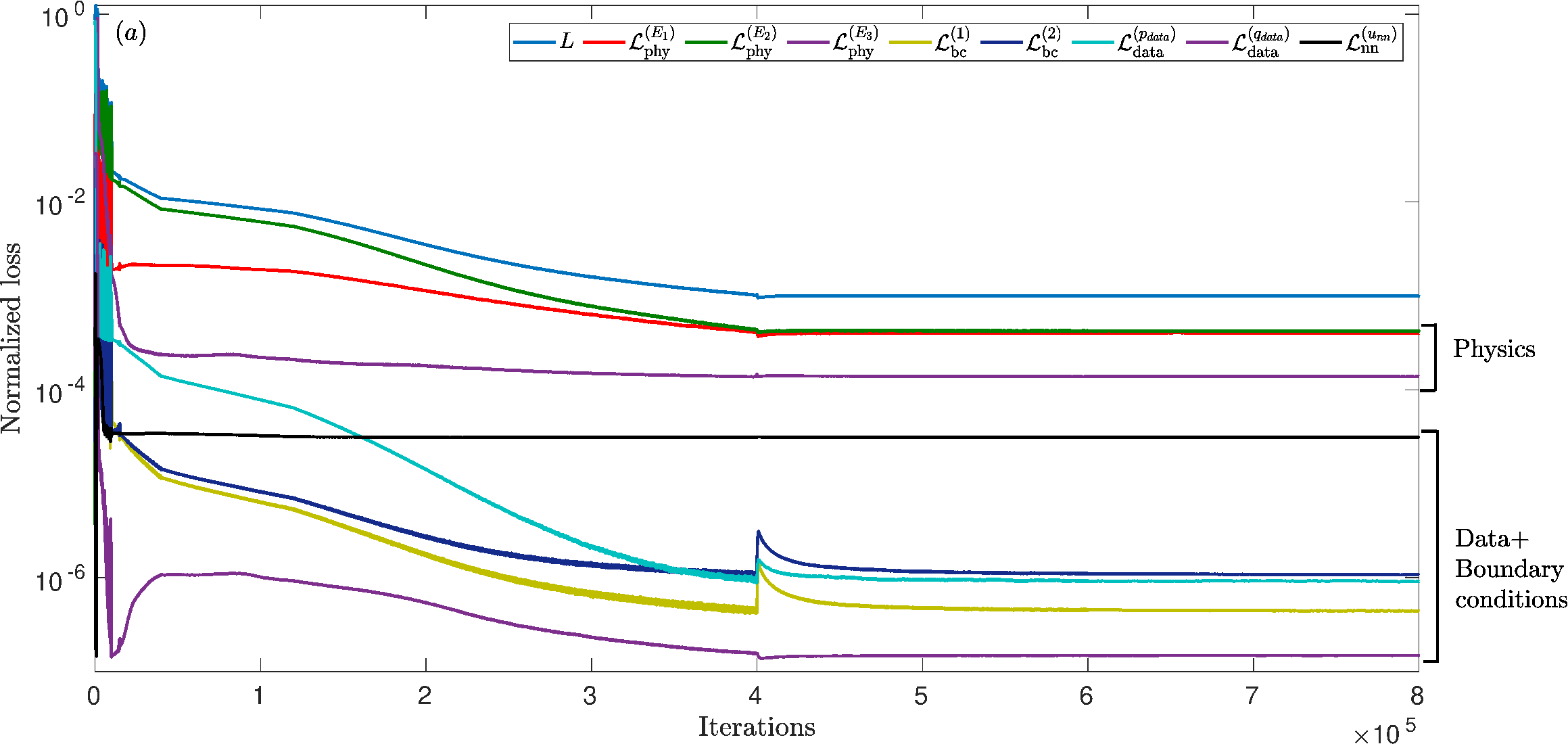}
\includegraphics[keepaspectratio,width=1\textwidth]{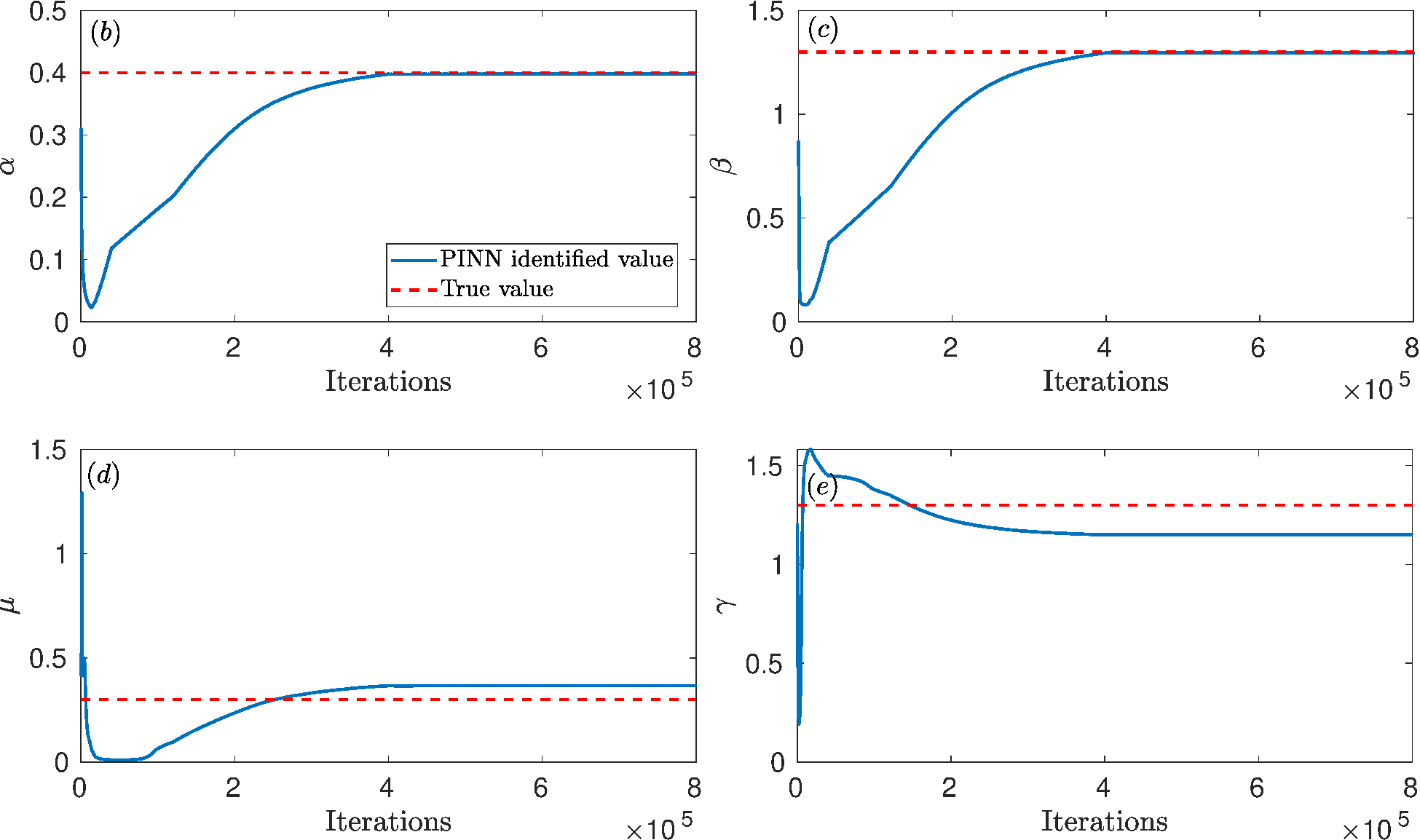}
\caption{\textbf{Case 4 - Minimization of the loss function and convergence of LOM parameters:} Panel $(a)$ shows the minimization of the various contributions to the loss function $\mathcal{L}$. Each contribution is normalized with $\mathcal{L}$ at the first iteration of the optimization. The last four panels $(b-e)$ illustrate the convergence of the parameters close to their true values. The figure is for a single realization. All the associated parameters and the test errors are listed in the third row of Table \ref{tab:a_lock_in_num}.}
\label{fig:loss_para_9}
\end{figure}

Figure \ref{fig:loss_para_9}$(a)$ shows the decrease of the various contributions to the loss $\mathcal{L}$ with the number of iterations. Each term is normalized with the value of $\mathcal{L}$ at the first iteration. Due to higher penalization on the data (acoustic pressure and heat release rate) and boundary conditions, their contributions decrease rapidly with the iteration. The decrease guides the decline of the physics loss to generate the entire $u,~p,~q_{pmt}$ fields. The subsequent panels ($b-e$) show the convergence of the parameters to their true values. In this realization, $\alpha$ and $\beta$ are predicted to their true values, while the other parameters have a slight offset. All the associated train and test errors are denoted in the third row of Table \ref{tab:a_lock_in_num}. The sharp jump at the iteration $4\times 10^5$ is due to the re-initialization of the self-adaptive weights to unity, thereby ensuring the global minimum in $\mathcal{L}$ is attained. 

\begin{figure}
\includegraphics[keepaspectratio,width=1\textwidth]{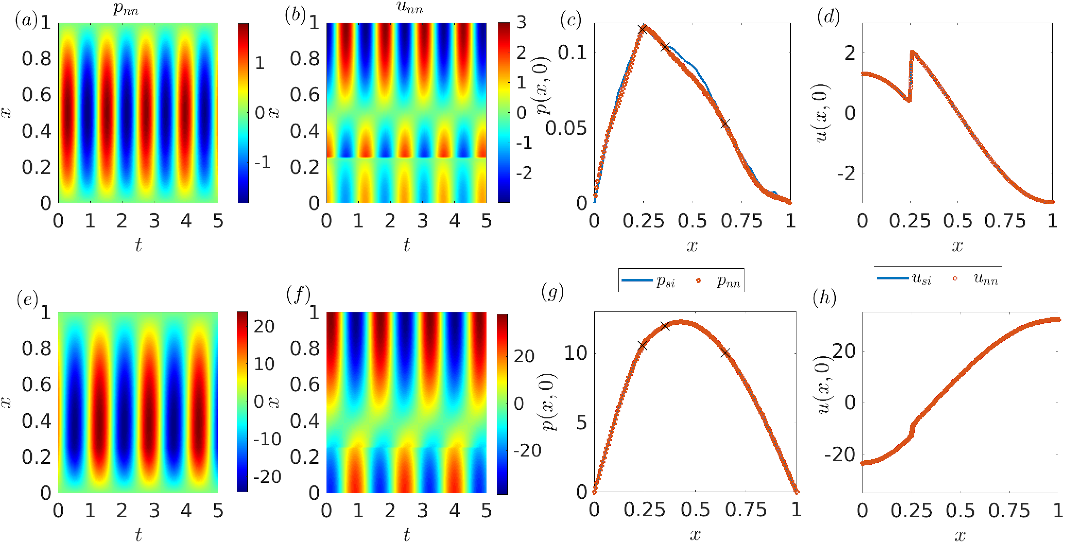}
\caption{\textbf{Case 4 - Study of PINN and simulated results in V- and A-lock-in regions:} Figure shows the performance of PINN to predict the entire acoustic field $u,~p$ of the combustor. The top and bottom rows correspond to a simulation in the V- and A-lock-in regions, respectively. The first two columns indicate the acoustic field. The next two columns compare the initial condition from the numerical and PINN solutions. Cross symbols in the third column indicate the locations where the acoustic pressure time series is fed to the PINN. The parameter values to the numerical solver and the predicted values from the PINN are given in the first row of Tables \ref{tab:v_lock_in_num} and \ref{tab:a_lock_in_num}, respectively.}
\label{fig:initial_condition_check_V_A_lock_i}
\end{figure}

As shown in figure \ref{fig:experimental_data_compilation}$(b)$, V- and A-lock-in regions have different acoustic oscillation amplitudes. Typical PINN results in both regions are described in figure \ref{fig:initial_condition_check_V_A_lock_i}. The top and bottom rows correspond to a simulation in the V- and A-lock-in regions, respectively. The first two columns indicate the acoustic field. There is an order of magnitude rise in the amplitude of the oscillations between V- and A-lock-in regions (refer to the contour levels of panels $a$, $b$, $e$, and $f$). A scaling factor of 3 is employed in PINN on the output variables ($p,~u$) to accurately capture the flow field. Scaling is performed only in the high-amplitude A-lock-in region. The same is applied in the subsequent cases. The following two columns compare the initial condition from the numerical and PINN solutions. A slight mismatch occurs for $p(x,0)$ (panel $c$) for the V-lock-in simulation. At the same time, a good match is observed for the rest of the variables (panels $d,~g-h$). It shows the ability of PINN to predict the initial condition in the entire spatial domain, which in many thermoacoustic experiments is unknown. Cross-symbols in the third column denote $p_{data}$ locations. 

The present neural network is trained only for 5 non-dimensional time units. On the other hand, experimental data are obtained for hundreds of non-dimensional time units. A time stitching process is, therefore, required. We use the ability of the PINN to reproduce the initial condition in the stitching process (described in \S \ref{subsec:stitching}).

Self-adaptive weights $\bm{\lambda_1,...\lambda_9}$, along with the mask function $\mathcal{F}$ is applied in the optimization process \eqref{eq:min_max_prob}. Figure \ref{fig:self_adap_weights_case_7}$(a,b)$ in the appendix shows the fields of the first two self-adaptive weights $\bm{\lambda_1,~\lambda_2}$, which corresponds to the physics loss of $p,~u$ (at the end of the last iteration). High values of $\bm{\lambda_1,~\lambda_2}$ indicate higher penalization at the locations of high gradients in the $p,~u$ field (figure \ref{fig:initial_condition_check_V_A_lock_i}$b$). As expected, the highest values of $\bm{\lambda_1,~\lambda_2}$ occur around the flame location (figure \ref{fig:self_adap_weights_case_7}$(a,b)$). Furthermore, the gradients in the acoustic wave pattern are captured in the field of self-adaptive weights. Other self-adaptive weights do exhibit similar behavior and, therefore, are omitted for conciseness. The bottom panel of figure \ref{fig:self_adap_weights_case_7} shows the averages (spatial and temporal) of the self-adaptive weights in the entire domain. Their increasing trend is in line with the optimization process \eqref{eq:min_max_prob}. 
 
Since the present case is the last one before deploying the experimental data, an extensive investigation of the PINN performance is performed in the two regions. Three sets of parameters in each region are selected. All the errors are collated in Tables \ref{tab:v_lock_in_num} (V-lock-in) and \ref{tab:a_lock_in_num} (A-lock-in). The parameters are chosen such that they fall in either V- or A-lock-in regions. In the V-lock-in region, we fix the frequency ratio $f=0.8$, while it is varied in the A-lock-in region for the three parameter sets. In this way, a variety of combinations can be examined.

The actual and predicted (mean and standard deviation) values of the parameters (first two columns) show a good match, if not for all the parameters. Furthermore, the training error of the data (third column in the tables) and the test error (last column in the tables) in $p,~u,~q_{pmt}$ shows a mean (over independent realizations) of less than 5\% less with standard deviation one order of magnitude lesser. We note that the mean test errors and their standard deviations in the current Tables (\ref{tab:v_lock_in_num} and \ref{tab:a_lock_in_num}) are lower than in Tables \ref{tab:PINN_drror} and \ref{tab:thermoacoustic_problem}, though the latter table deals with the simpler problem: cases 1-3. Higher accuracy is therefore attributed to the grid clustering around the flame location, allowing $q_{pmt}$ to significantly contribute to the physics loss of $E_2$. 

Simulations are performed with acoustically open boundary conditions. Therefore, the loss weights are kept higher than the governing equations. On the other hand, in the experiments, the boundary conditions are unknown. We could not obtain a converged solution non-zero solution for the parameters when their loss weights $a_{bc_1},~a_{bc_2}$ are set to zero. We do, however, know that the acoustic pressure boundary conditions are close to zero. Therefore, a lower value of the loss weight $a_{bc_1}=a_{bc_2}=0.1$ is a possible trade-off when applied to the experiments. 

At the same time, we reduced the value of the loss weight $a_{p_{data}}$. This allows the hybrid PINN solution to be optimum for both the physics and the data parts. A Tikonov regularizer is additionally required to obtain a converged non-zero solution for the parameters. A value of $10^{-4}$ is chosen for the regularization coefficient. A sub-case, termed case 4a, is performed with the simulated data to understand the performance of PINN. 

The last two rows of Table \ref{tab:a_lock_in_num} show the comparison between cases 4 and 4a for the same simulated data. The mean value of the parameters is predicted closely between the two cases, although there is an increase in the standard deviation. The same trend is observed in the training and test errors, too (last two columns). The increase in the standard deviation is expected due to the following reason. The two boundary conditions act as additional information for case 4. Therefore, a lower penalization of them in $\mathcal{L}$ in case 4a leads to a rise in the mean and standard deviation of the predicted parameters and the training/test errors from the simulated solution. In any case, the standard deviation is at least one order less than the mean in case 4a. Hence, we believe that the network architecture of case 4a is as good as case 4. The above test also emphasizes measuring acoustic pressure near the ends of the combustor in future experiments. 

Overall, the above discussion indicates that the neural network architecture and the weights in the loss function serve to capture the hidden flow field in the V- and A- lock-in regions accurately. The network is ready to deal with the actual experimental data with the loss weights given in case 4a.     

\section{Experimental results: case 5 - PINN solution and parameter identification using the experimental data}\label{sec:inv_prob_exp}
The section is divided into three subsections. In the first one, we discuss the preprocessing of the experimental data for the PINN to provide non-trivial results. In the second subsection, we deal with the evaluation of PINN results, followed by temporal stitching to construct the entire time series. In the last subsection, we provide physical interpretations based on the probability distribution of the LOM parameters obtained for different time segments and slpm.

\subsection{Pre-processing the turbulent heat release rate}
\begin{figure}
\includegraphics[keepaspectratio,width=1\textwidth]{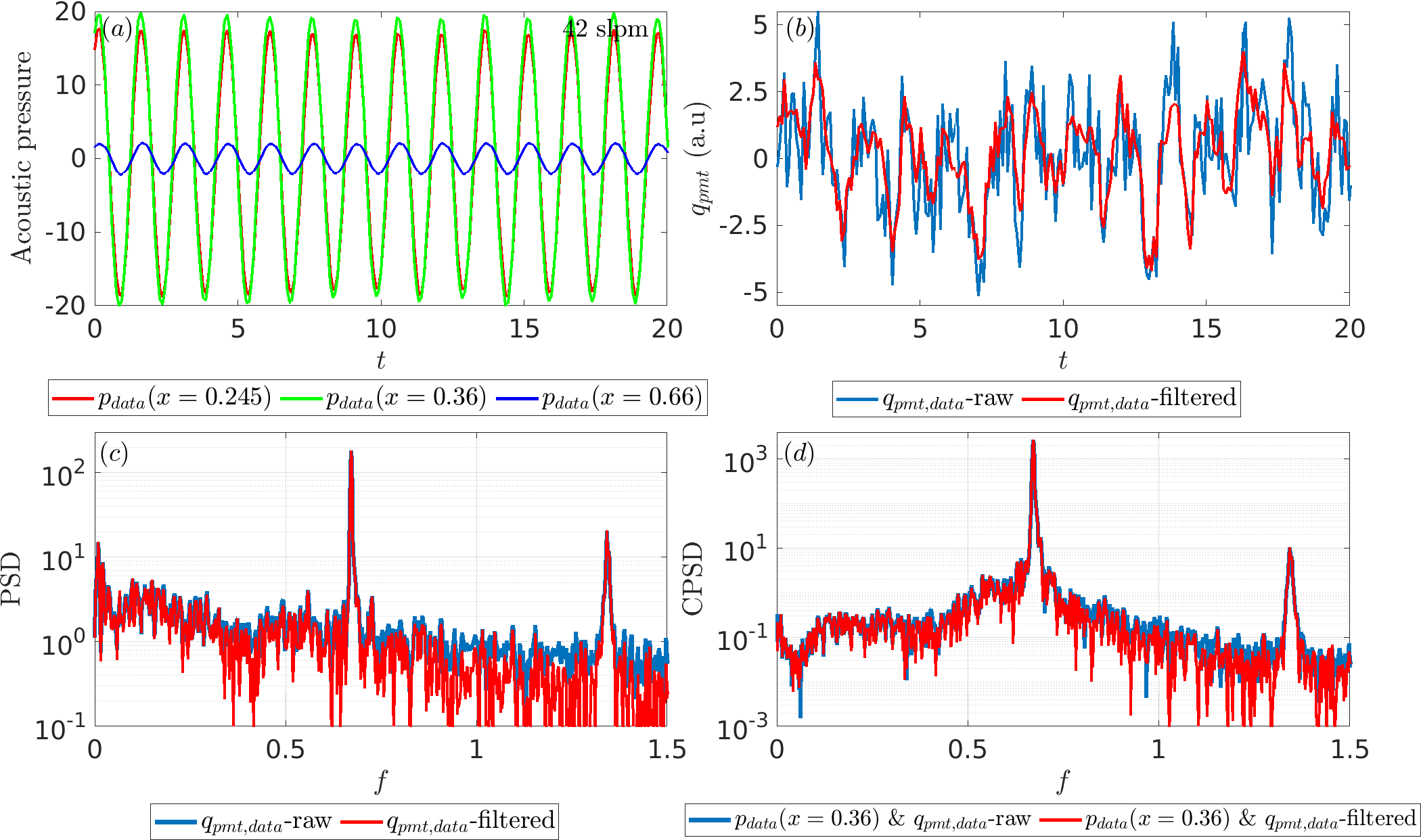}
\caption{\textbf{Case 5 - Preprocessing of the experimental data:} The top panels show the evolution of the experimental non-dimensionalized $(a)$ acoustic pressure and $(b)$ unsteady heat release rate fluctuations. The bottom panels show their corresponding power spectral density (PSD). Blue and red curves in panels $(b-d)$ indicate the raw and filtered $q_{pmt,data}$. The airflow rate is 42 slpm.}
\label{fig:experimental_pre_processing}
\end{figure}
The flow in the combustor is turbulent (Reynolds number based on the diameter of the bluff body $\sim 10^4$). Figure \ref{fig:experimental_pre_processing} shows a snippet of the non-dimensionalized acoustic pressure (panel $a$) and unsteady heat release rate (panel $b$). Even during the large amplitude oscillation (A-lock-in), the raw $q_{pmt,data}$ (blue curve) is highly turbulent, while the acoustic pressure follows a well-defined pattern. Although the former is turbulent, we observe well-defined peaks (blue curve) both in power (PSD) and cross-power spectral densities (CPSD) in the bottom panels. The sharp peaks of CPSD occurring at the same frequencies as the PSD indicate a strong thermoacoustic coupling. 

Although the turbulent fluctuations in $q_{pmt,data}$ are present, we avoided modeling them in LOM with an additional stochastic term for simplicity. 
We could not obtain a converged non-trivial solution for the parameters and the variables ($p,~u,~q_{pmt}$). Therefore, we decided to Fourier filter the signal $q_{pmt,data}$, and remove the Fourier frequencies, whose amplitudes are less than 5\% of the dominant amplitude. The above cut-off served as a trade-off between the convergence of the non-trivial PINN solution and the drop in the root mean square value of the filtered signal. The $q_{pmt,data}$ filtered PSD and CPSD (red curve) in the bottom panels closely match the dominant Fourier components, indicating no significant drop in the amplitudes. On the other hand, the filtering process allows us to see patterns in $q_{pmt,data}$ (red curve in panel $b$) both visually and with the convergence of the PINN solution. Therefore, a 5\% filtering preprocessing step is performed only in $q_{pmt,data}$. 

The ``pure" acoustic frequency $(\tilde f_a)$ of the combustor measured from the experiment is 230.5 Hz. The word ``pure" indicates the frequency of the combustor without the interaction between the combustor acoustic field and vortex shedding. ``Pure" vortex shedding frequency ($\tilde f_v$ from figure 6b of \citealt{singh2021experimental}) for various flow rates is used. Cubic extrapolation is performed for flow rates where the pure vortex shedding frequency is unavailable (A-lock-in region). One obtains the frequency ratio $f=\tilde f_a/\tilde f_v$. In the non-dimensional form, solving \eqref{eq:disp_eq} provides the pure acoustic frequency $f_a$. Knowing $f$ from the experiments allows us to calculate $\omega_v$. A stitching process is required for the experimental data once the PINN solution is obtained.

\subsection{PINN solution and stitching}\label{subsec:stitching}
\begin{figure}
\includegraphics[keepaspectratio,width=1\textwidth]{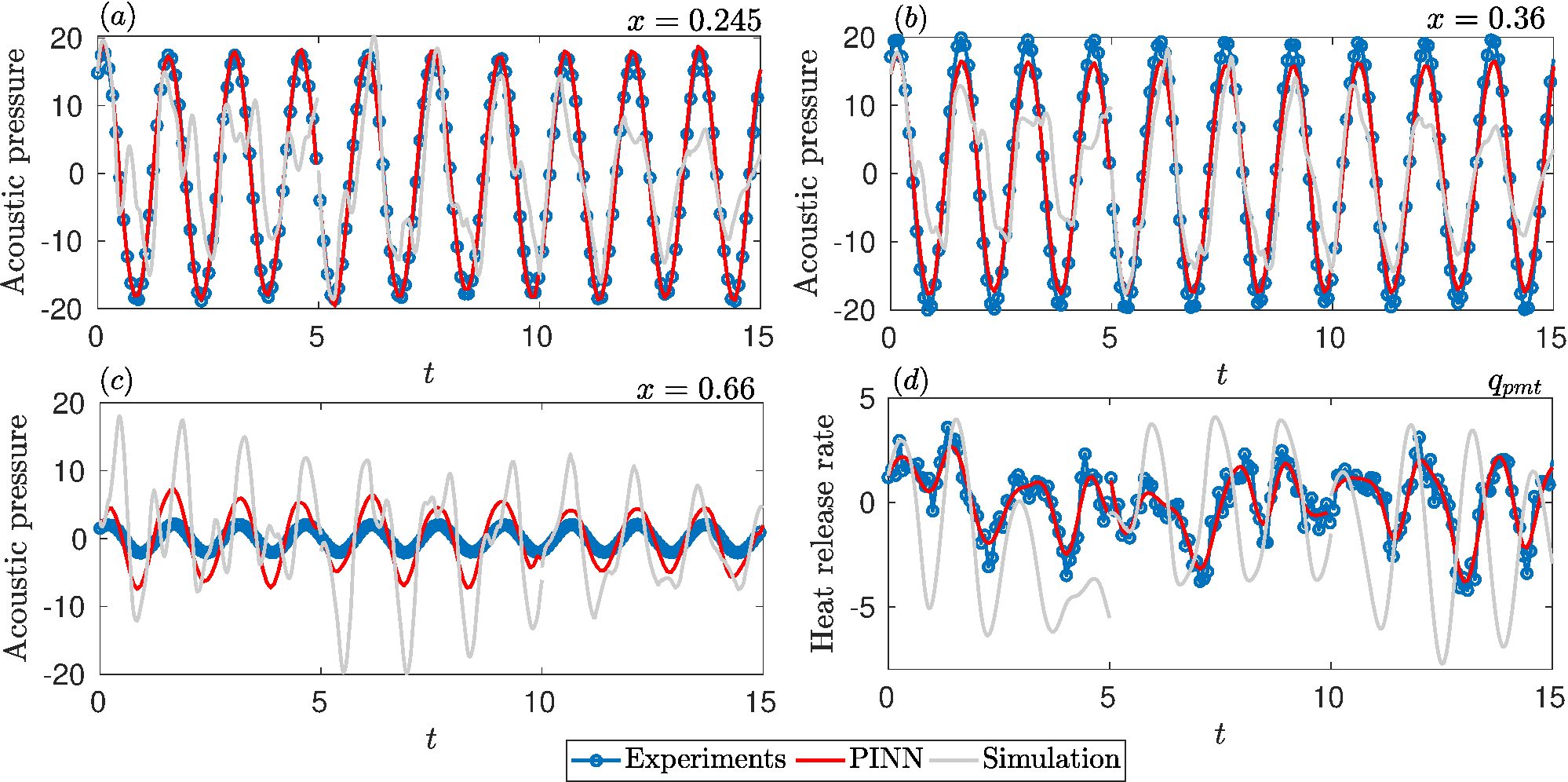}
\includegraphics[keepaspectratio,width=1\textwidth]{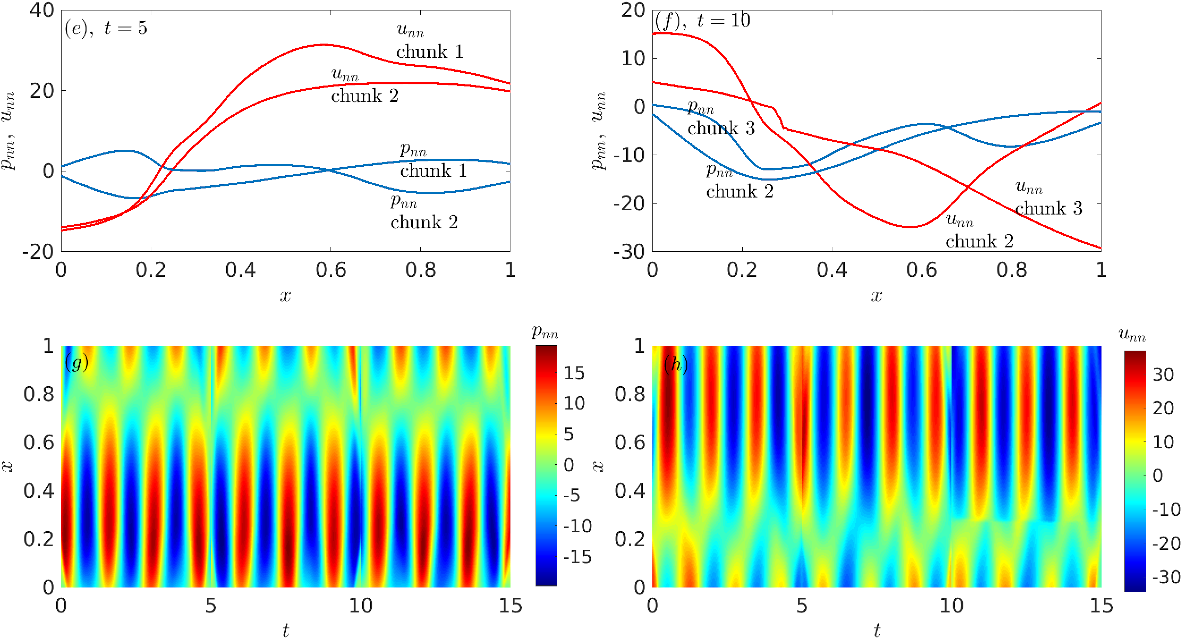}
\caption{\textbf{Case 5 - PINN solution from the experimental data:} Panels $a-d$ compares the evolution of acoustic pressure and heat release measured from the experiments, predicted by PINN and simulation. Panels $(e-f)$ compare the spatial distribution of pressure and velocity at the chunk intersections $t=5$ and $10$, respectively. The lowest panel $(g-h)$ shows the entire evolution of $p_{nn}$ (panel $g$) and $u_{nn}$ (panel $h$), respectively. The airflow rate is 42 slpm.}
\label{fig:42_slpm_stitching}
\end{figure}
The neural network is trained for each 5 non-dimensional time chunks. Therefore, the experimental data is cut into chunks spanning 5 non-dimensional time units. The parameters and the flow field are obtained for each chunk. A total of 50 chunks (250 non-dimensional time units) are used for the analysis. Figure \ref{fig:42_slpm_stitching} shows the results for three consecutive chunks. Acoustic pressure at the three measurement locations and the heat release rate are shown in the first four panels $(a-d)$. The open blue circle and solid red line correspond to the experimental data and the PINN solution. The large amplitude pressure oscillations (top panels) are captured well. The error is relatively large at the low amplitude acoustic pressure oscillations at $x=0.66$ (panel $c$). This may be acceptable in a design process as such regions in a combustion chamber do not contribute much to the fatigue loads to the combustor. Surprisingly, the time-snippet of the turbulent heat release rate is captured well by the PINN (panel $d$), in spite of the simple LOM usage. Importantly, in all the panels $(a-d)$, the phase of the oscillation is captured extremely well. As is the fact that the phase difference between heat release and pressure fluctuations determines the driving and damping of a thermoacoustic oscillation (Rayleigh criteria: \citealt{rayleigh78}).   

The gray solid curve in the first four panels represents the results of the LOM. For every chunk, we use the identified parameters and initial conditions from the PINN to perform the simulation. LOM performs poorly in comparison to the PINN solution in capturing the experimental time series. A discontinuity at the chunk intersections ($t=5,10,15...$) occurs as the neural network for PINN is trained independently for each chunk. The solution predicted for acoustic pressure (blue curve) and velocity (red curve) by PINN at the first (panel $e$) and second (panel $f$) chunk intersections only matches the trend. It occurs as the chunks are trained independently. A future work can include the jump at the intersections as an additional loss term in $\mathcal{L}$ as done in \cite{Jagtap_2020_XPINN}. The PINN solution does not suffer from the discontinuity problem, at least at the measurement locations, and reproduces the observed experimental time series fairly well. Temporal discontinuities in the acoustic field (other than the measurement locations) do appear in the PINN solution at the time chunk intersections (panels $g,h$). They are faint, and the temporal locations of acoustic pressure and velocity nodes and anti-nodes are not much affected due to the stitching process. Therefore, they provide an impression of almost temporally continuous flow fields, as one expects them to be.     

The evolution of PINN pressure $p_{nn}$ (panel $g$) and velocity $u_{nn}$ (panel $h$) can thus be considered as a good representation of the actual fields existing in a combustor. Despite the discontinuities, the spatial distribution of the magnitudes of $p_{nn}$ and $u_{nn}$ serve as an important input for fatigue and thermal load calculations, respectively. They serve as design inputs for structural reinforcements and thermal protection systems of gas-turbine and rocket engines \citep{culick88}.
\begin{figure}
\centerline{\includegraphics[keepaspectratio,width=1\textwidth]{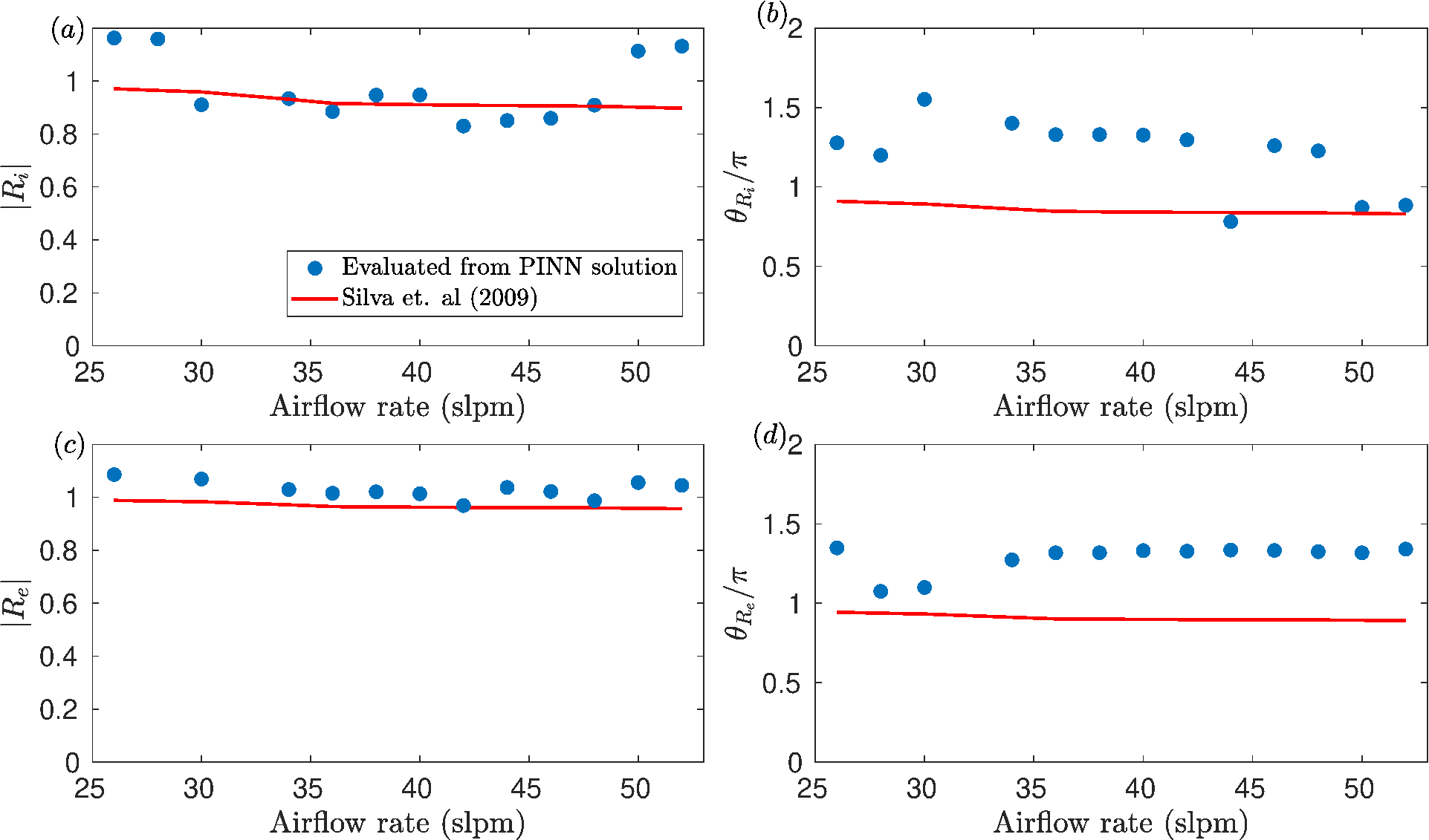}}
\caption{\textbf{Case 5 - Study of reflection coefficient predicted from PINN and a literature model:} Comparison of the PINN predicted absolute $(a,~c)$ and phase $(b,~d)$ of the inlet ($R_i$ - top panels) and exit ($R_e$ - bottom panels) reflection coefficients of the combustor with \cite{silva2009approximation}.}
\label{fig:reflection_coefficient_comparison}
\end{figure}

As shown in the previous cases (\S \ref{sec:sol_acoustic_eq}-\ref{sec:inv_prob_syn}), we require to run multiple realizations for each chunk to obtain the mean and standard deviation of the predicted model parameters and flow field. For chunk 1 ($t=0-5$), the predicted LOM parameters are $\alpha=0.23\pm0.02,~\beta=0.1\pm 0.02,~\mu=0.13\pm 0.05,~\gamma=5.3\pm 0.2$. The train errors for $p_{data}$ and $q_{data}$ are $19\pm7.6$\% and $33\pm4.3$\%, respectively. Figure \ref{fig:42_slpm_multiple_runs_std} in the appendix shows the mean field of the variables ($p,~u,~q_{pmt}$) and standard deviation over the realizations. Standard deviation is one order lesser than the corresponding mean value throughout the entire field, which indicates a small epistemic uncertainty. Furthermore, running ten realizations for 50 time chunks and all the flow rates leads to a high computational cost. In view of the above two points, we restrict ourselves to one realization in each chunk for the experimental data. 

Although PINN predicts the hidden variable acoustic velocity, its value cannot be verified directly as its experimental data is not measured at any location. In order to have confidence in the results, we calculate the reflection coefficients at the entry $(R_i)$ and exit $(R_e)$ of the combustor from the PINN solution at the dominant frequency (figure \ref{fig:reflection_coefficient_comparison}). The acoustic velocity is required to calculate the reflection coefficient. The results are compared with model 1 (flanged-end) of \cite{silva2009approximation}. A decent comparison in the reflection coefficient, both in its absolute $(|R_i|,~|R_e|)$ and phase $(\theta_{R_i},~\theta_{R_e})$, is observed. We expect the mismatch due to the (i) difference in the actual geometry at the combustor ends and the flanged-end approximation and (ii) missing physics in the LOM. In any case, the decent comparison indicates a belief in the predicted acoustic velocity field. In the following section, we provide a physical interpretation based on the variation of the LOM parameters.

\subsection{Physical interpretation}\label{subsec:phy_interp}
\begin{figure}
\includegraphics[keepaspectratio,width=1\textwidth]{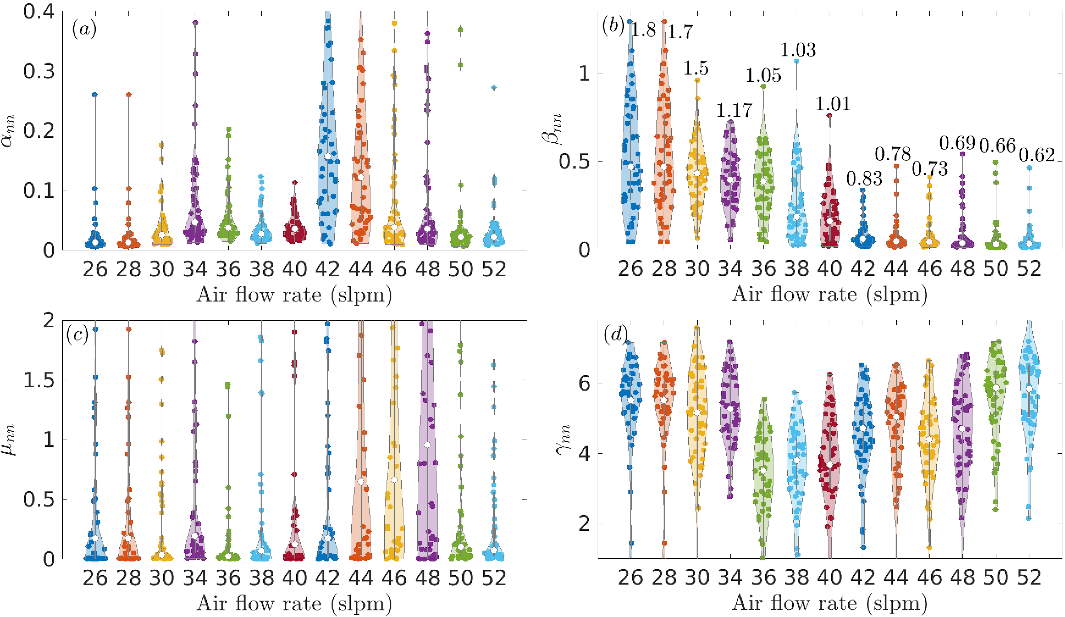}
\includegraphics[keepaspectratio,width=1\textwidth]{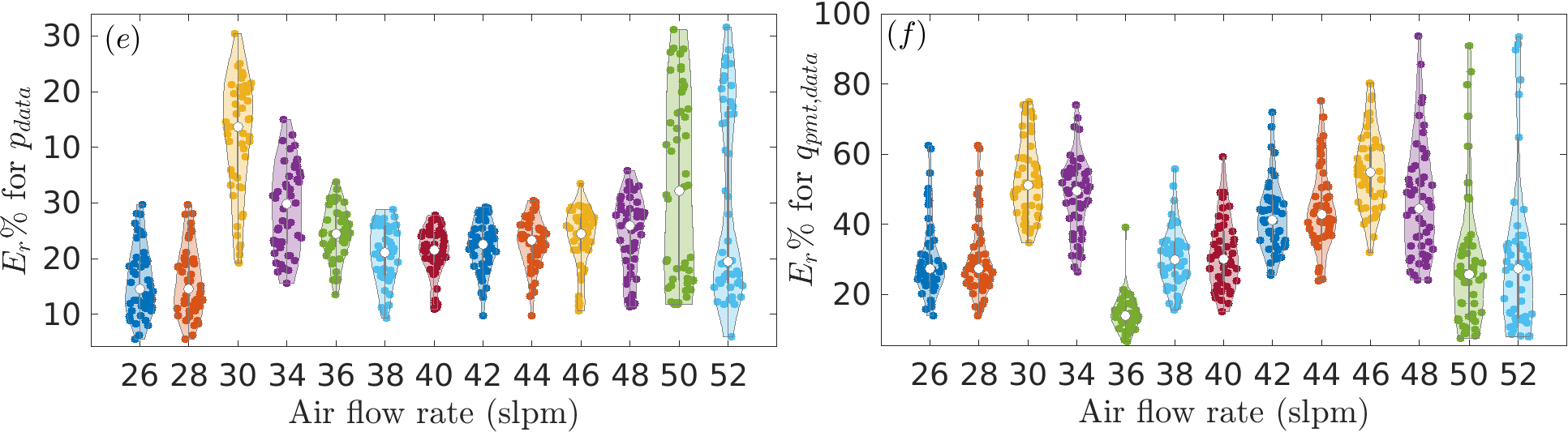}
\caption{\textbf{Case 5 - Probability density functions of the LOM parameters and train errors:} Violin plots showing the probability density functions (estimated using kernel density estimation) for all the PINN predicted parameters (panels $a-d$) and 
 train errors associated with $p_{data}$ (panel $e$), $q_{data}$ (panel $f$).}
\label{fig:violin_plot_parameter_data_error}
\end{figure}

Figure \ref{fig:violin_plot_parameter_data_error} shows the violin plots containing the probability density functions (PDFs) of all the parameters and the train errors for the entire time chunk at various airflow rates. The (PDFs) are estimated from the histograms using a kernel density estimator. The kernel used is a sum of a number of standard normal distributions, which fits the histogram optimally in a least-square sense. The data used in the histogram are shown as dots. In the violin plots, the PDFs are vertically symmetric about the center gray line. It is performed to distribute the dots on either side equally, allowing a better visualization. We observe that the PDFs are highly non-Gaussian. Therefore, the median (represented as a hollow white circle) over the mean is preferred in the subsequent physical interpretations. Furthermore, thin regions of the PDFs are caused by either an outlier or a low tail of the PDFs. They do not significantly alter the median. In order to show the region around the median, the thin regions in some plots (for example, panel $c$) are trimmed. The frequency ratio $f$ is also marked in panel $(b)$ to aid our interpretation.

Figure \ref{fig:violin_plot_parameter_data_error} is interpreted in relation to figure \ref{fig:experimental_data_compilation}. Panel $(a)$ of the latter figure indicates the boundary between V- and A-lock-in regions occurs around 40 slpm ($f=1.01$). We make the following interpretation based on the variation of the LOM parameter values. 
\begin{itemize}[label=$\blacksquare\;\;$, leftmargin=0.5cm, itemindent=0pt]
\setlength{\itemsep}{6pt}
    \item The median value of $\beta_{nn}$ (panel $b$) dictates the influence of the unsteady heat release rate on the acoustic field. It is high in the V-lock-in region, indicating that the unsteady heat release rate caused by ``natural" vortex shedding drives the acoustic field. $\beta_{nn}$ sharply decreases near the boundary of the A-and V-lock-in regions, indicating that the natural vortex shedding is no longer strong enough to drive the acoustic field.
    \item On the other hand, $\gamma_{nn}$ (panel $d$) shows an almost gradual rise in its median for flow rates above 36 slpm. This indicates that the acoustic field begins to influence the vortex-shedding process, leading to a strong thermoacoustic interaction, i.e., the A-lock-in region. We observed it as a sharp rise in the root mean square value of the acoustic pressure (figure \ref{fig:experimental_data_compilation}$b$). We also observe high values of $\gamma_{nn}$ for lower airflow rates ($<34$ slpm), where A-lock-in does not occur. We hypothesize that A-lock-in did not happen as (i) $f$ is far away from 1 and (ii) lock-in region is not symmetrical about $f=1$ as observed in some experimental \citep{li13c}/numerical \citep{britto2019lock} investigations. 
    \item The nonlinearity $\mu_{nn}$ (panel $c$) presents a similar trend to the acoustic pressure amplitude (figure \ref{fig:experimental_data_compilation}$b$). It is in line with the well-known feature of nonlinearity kicking in at higher amplitudes \citep{strogatz00}.
    \item The last parameter $\alpha_{nn}$ (panel $a$) has high values at 42-44 slpm, where the pressure and, therefore, the velocity amplitudes clock the highest values (figure \ref{fig:experimental_data_compilation}$b$). We hypothesize that large amplitudes trigger the boundary layer acoustic dissipation inside the combustor, leading to the observed high values in $\alpha_{nn}$.
    \item The median training error $E_r$ of the pressure data hovers around 20\%. A higher median $E_r$ for the heat release data with a large spread indicates the inability of the van der Pol oscillator to represent the turbulent flame.
\end{itemize}
\vspace{0.5cm}
The above interpretations are made with respect to the trends of the LOM parameters with the airflow rate. These interpretations can be expected heuristically from the form of the LOM (\ref{eq:final_damped_wave_eq_temp_jump_mom}-\ref{eq:van_der_pol_osc}). However, LOM's numerical values allowed us to generate a quantitatively accurate hybrid PINN model. Such a model can serve as a design input to develop thermoacoustically stable and structurally efficient combustors.

\section{Concluding remarks and future directions}\label{sec:conc}
The physics-informed neural network (PINN) is employed to obtain a hybrid neural network-based solution for a thermoacoustic system. PINN optimally blends the experimental data and low-order model (LOM). PINN is applied to study thermoacoustic interactions in a combustor with a flame anchored by a bluff body, shedding vortices. The interaction displays the feature of lock-in, where the shedding frequency is either close to the vortex shedding (V-lock-in) or acoustic (A-lock-in) of the combustor. A large amplitude combustion instability is observed during A-lock-in. 

The major objective is to apply PINN to existing experimental data, which was not presumed to have its application in machine learning at the time of the measurement. Such is the case for a vast majority of the experimental data on thermoacoustic instability available in the literature. The problem is challenging as it is mathematically ill-posed (no initial or boundary conditions), and experimental data from sparse spatial locations (similar to the practical systems) are available. Acoustic equations and a van der Pol oscillator are employed as the LOM. Acoustic pressure at three locations and the unsteady heat release rate of the flame are used as the data. Combining the above two, PINN is able to perform the following. (i) Development of a hybrid neural network model for the acoustic field and unsteady heat release rate. (ii) Construction of the entire acoustic pressure and velocity fields. The latter is a challenging quantity to measure in the entire field of combustors. (iii) Estimation of the LOM parameters. The variation of their values with the airflow rates illustrated physical interpretations. On the whole, PINN is a promising framework to reconstruct the unknown flow fields in a combustor and possibly bring out the hidden physics of a variety of past and the design of new combustors. In future work we will attempt to improve the LOM by identifying the missing physics \citep{zhao2021physics,zou2023correcting}, employ Bayesian-PINN for providing a rigorous uncertainty quantification for both parameters and the states \citep{yang2021b,li_2023surrogate}, and reduce the computational cost by using methods such as transfer learning \citep{prantikos2023physics,xu2023transfer}.   

\section*{Declaration of interests} The authors report no conflict of interest.

\section*{Acknowledgement}\label{sec:ack}
The first author thanks the Fulbright-Nehru Academic and Professional Excellence fellowship, award number 2734/F-N APE/2022, and the sabbatical from his institute for funding his stay at Brown University, USA. Partial financial support from the Science and Engineering Research Board, India, through the grant YSS/2015/000351 is greatly acknowledged. The work of K. Nath and G. E. Karniadakis was supported by OSD/AFOSR MURI grant FA9550-20-1-0358. All the computations are performed using the resources of the Center for Computation and Visualization, Brown University. 

\bibliographystyle{jfm}
\bibliography{sathesh_literature}\label{refs} 

\clearpage
\appendix
\renewcommand{\thesection}{\Alph{section}}
\renewcommand{\thesubsection}{\Alph{section}.\arabic{subsection}}
\renewcommand{\thesubsubsection}{\Alph{section}.\arabic{subsection}.\arabic{subsection}}

\setcounter{equation}{0}
\renewcommand{\theequation}{\thesection.\arabic{equation}}
\setcounter{table}{0}
\renewcommand{\thetable}{\Alph{section}\arabic{table}}
\setcounter{figure}{0}
\renewcommand{\thefigure}{\Alph{section}\arabic{figure}}
\setlength\belowcaptionskip{20pt}
\section{Appendix A: Additional table for hyperparameters and test errors} 
\label{app:hyperparameters}
\begin{center}
\begin{tabular}{ccccccc}
\cline{1-7}
Iterations    & $lr_{\bm{\theta}},lr_{\bm{\varLambda}}$ & $lr_{\bm{\lambda}_1},lr_{\bm{\lambda}_2},lr_{\bm{\lambda}_3}$ & $lr_{\bm{\lambda}_4},lr_{\bm{\lambda}_5}$ & $lr_{\bm{\lambda}_6},lr_{\bm{\lambda}_7}$ & $lr_{\bm{\lambda}_8}$   & $lr_{\bm{\lambda}_9}$ \\ \cline{1-7}
1-10000       & $0.5\times 10^{-3}$           & $1\times10^{-2}$                               & $2.5\times10^{-2}$              & $5\times10^{-2}$                & $2.5\times10^{-2}$ & $1\times10^{-2}$ \\ \cline{1-7}
10001-15000   & $0.5\times10^{-4}$            & $1\times10^{-2}$                               & $2.5\times10^{-2}$              & $5\times10^{-2}$                & $2.5\times10^{-2}$ & $1\times10^{-2}$ \\ \cline{1-7}
15001-40000   & $0.5\times10^{-5}$            & $1\times10^{-2}$                               & $2.5\times10^{-2}$              & $5\times10^{-2}$                & $2.5\times10^{-2}$ & $1\times10^{-2}$ \\ \cline{1-7}
40001-120000  & $0.5\times10^{-6}$            & $1\times10^{-2}$                               & $2.5\times10^{-2}$              & $5\times10^{-2}$                & $2.5\times10^{-2}$ & $1\times10^{-2}$ \\ \cline{1-7}
120001-150000 & $1\times10^{-7}$              & $1\times10^{-2}$                               & $2.5\times10^{-2}$              & $5\times10^{-2}$                & $2.5\times10^{-2}$ & $1\times10^{-2}$ \\ \cline{1-7}
150001-600000 & $1\times10^{-8}$              & $1\times10^{-2}$                               & $2.5\times10^{-2}$              & $5\times10^{-2}$                & $2.5\times10^{-2}$ & $1\times10^{-2}$ \\ \cline{1-7}
600001-800000 & $1\times10^{-9}$              & $1\times10^{-2}$                               & $2.5\times10^{-2}$              & $5\times10^{-2}$                & $2.5\times10^{-2}$ & $1\times10^{-2}$ \\ \cline{1-7}
\end{tabular}
\captionof{table}{\textbf{Learning rate scheduling:} Learning rate scheduling for weights, biases of the neural network and self-adaptive weights used in the optimization of loss function $\mathcal{L}$ (\ref{eq:opt_proc_1}-\ref{eq:opt_proc_2}).}\label{tab:lr_sch}
\end{center}

\begin{center}
\begin{tabular}{C{1cm}C{0.75cm}C{0.75cm}C{0.75cm}C{0.75cm}C{0.75cm}C{0.75cm}C{0.75cm}C{0.75cm}C{0.75cm}C{0.75cm}}
\cline{1-11}
Case & $a_{E_1}$ & $a_{E_2}$ & $a_{E_3}$ & $a_{bc_1}$ & $a_{bc_2}$ & $a_{ic_1}$ & $a_{ic_2}$ & $a_{p_{data}}$ & $a_{q_{data}}$ & $a_{u_{nn}}$ \\ \cline{1-11}
1    & 1         & 1         & 0         & 1          & 1          & 1          & 1          & 0              & 0              & 0              \\ \cline{1-11}
2    & 1         & 1         & 0         & 1          & 1          & 0          & 0          & 1              &        1       & 0.1            \\ \cline{1-11}
3    & 1         & 1         & 1         & 1          & 1          & 0          & 0          & 10             & 10             & 0.1            \\ \cline{1-11}
4    & 1         & 1         & 1         & 10         & 10         & 0          & 0          & 10             & 50             & 0.1            \\ \cline{1-11}
4a   & 1         & 1         & 1         & 0.1        & 0.1        & 0          & 0          & 5              & 50             & 0.1            \\ \cline{1-11}
5    & 1         & 1         & 1         & 0.1        & 0.1        & 0          & 0          & 5              & 50             & 0.1            \\ \cline{1-11}
\end{tabular}\captionof{table}{\textbf{Loss weights:} Table showing the loss weights in the loss function $\mathcal{L}$ \eqref{eq:loss_fn_gen} for each case.}\label{tab:weights_case}
\end{center}

\begin{center}
\begin{tabular}{C{1cm}L{3cm}C{0.5cm}C{2.5cm}C{2.5cm}C{2.5cm}}
\hline
\multirow{2}{*}{\#} & \multirow{2}{*}{Acoustic mode} & \multirow{2}{*}{$\alpha$} & \multirow{2}{*}{$\omega/2\pi$}& \multicolumn{2}{c}{Test error $E_r$ \%} \\ \cline{5-6} 
&  & &  & $p_{nn}$ & $u_{nn}$  \\ \cline{1-6}
\multirow{4}{*}{Case 1} & Fundamental & 0.2 & $0.66 + 0.016i$ & $1.8\pm 0.26$ & $2.1\pm 0.26$   \\ \cline{2-6}
& Fundamental & 0.3      & $0.66 + 0.024i$                                                              & $2.0\pm 0.22$ & $2.6\pm0.22$                        \\ \cline{2-6}
& \begin{tabular}[c]{@{}l@{}}Fundamental\\ +first harmonic\end{tabular}                         & 0.2      & \begin{tabular}[c]{@{}l@{}}$0.66 + 0.016i$\\ $1.46 + 0.016i$\end{tabular} & $3.7\pm 0.38$ & $4.1\pm 0.44$                        \\ \cline{2-6}
& \begin{tabular}[c]{@{}l@{}}Fundamental\\ +first harmonic\end{tabular}                         & 0.3      & $0.66 + 0.024i$                                                              & $3.8\pm 0.5$ & $4.4\pm 0.5$                        \\ \cline{1-6}
\multirow{2}{*}{Case 2}  & Fundamental, with synthetic data & 0.2  & $0.66 + 0.016i$  & $2.1\pm 9.5\times 10^{-2}$ & $4.5\pm3.1\times 10^{-2}$                        \\ \cline{2-6}
& \begin{tabular}[c]{@{}l@{}}Fundamental + first \\ harmonic, with \\ synthetic data\end{tabular}  & 0.2  & \begin{tabular}[c]{@{}l@{}}$0.66 + 0.016i$\\ $1.46 + 0.016i$\end{tabular} & $2.2\pm 6.6\times 10^{-2}$ & $4.6\pm2.5\times 10^{-2}$                        \\ \cline{1-6}
\end{tabular}
\captionof{table}{\textbf{Test error $E_r$ for cases 1-2 (\S \ref{sec:sol_acoustic_eq}-\ref{sec:synthetic_data}):} Mean and standard deviation of the test error $(E_r)$ in the PINN solution for (i) two values of the damping coefficient $\alpha$, (ii) purely fundamental and (iii) combined fundamental and first harmonic of the acoustic modes. The ratio of the Fourier amplitude of the first harmonic and the fundamental mode is 0.07.}\label{tab:PINN_drror}

\begin{tabular}{C{2cm}L{4cm}L{4cm}}
\cline{1-3}
LOM parameters & \multicolumn{1}{l}{Train error $E_r$ \%} & \multicolumn{1}{l}{Test error $E_r$ \%}  \\ \cline{1-3}
\begin{tabular}[c]{@{}l@{}}$\alpha=0.2$\\ $\beta=1.1$\\  $\mu=0.2$\\ $\gamma=1.3$\end{tabular}                                                 & \begin{tabular}[c]{@{}l@{}}$p:~ 2.2\pm 7.7\times 10^{-2}$\\ $q_{pmt}:~ 0.86\pm 4.9\times 10^{-2}$\end{tabular} & \begin{tabular}[c]{@{}l@{}}$p:~2.4\pm 0.10$\\ $u:~5.7\pm 0.70$\\ $q_{pmt}:~1.2\pm 8.2\times 10^{-2}$\end{tabular} \\ \cline{1-3}

\begin{tabular}[c]{@{}l@{}}$\alpha=0.2$\\ $\beta=1.1$\\  $\mu=0.2$\\ $\gamma=0.3$\end{tabular}                                                 & \begin{tabular}[c]{@{}l@{}}$p:~ 0.48\pm 4.1\times 10^{-2}$\\ $q_{pmt}:~ 1.2\pm 5.5\times 10^{-2}$\end{tabular} & \begin{tabular}[c]{@{}l@{}}$p:~1.4\pm 0.13$\\ $u:~5.7\pm 0.67$\\ $q_{pmt}:~3.8\pm 0.82$\end{tabular} \\ \cline{1-3}
\end{tabular}\captionof{table}{\textbf{Train and test errors $E_r$ for case 3 (\S \ref{sec:heaT_source_inclusion}):} Table showing the known LOM parameters (first column), along with the train (second column) and test (third column) errors in the PINN solution. The frequency ratio $f=1.1$.}\label{tab:thermoacoustic_problem}
\end{center}

\begin{center}
\begin{tabular}{llll}
\cline{1-4}
True value                                                                                                                                  & PINN predicted value                                                                                                                                                                          & Train error $E_r$ \%                                                                                         & Test error $E_r$ \%                                                                                                                    \\ 
\cline{1-4}
\begin{tabular}[c]{@{}l@{}}$\alpha=0.2$\\ $\beta=1.7$\\  $\mu=0.3$\\ $\gamma=1.3$\end{tabular}                                               & \begin{tabular}[c]{@{}l@{}}$\alpha=0.2\pm 6.8\times 10^{-3}$\\ $\beta=1.7\pm 5.3\times 10^{-2}$\\ $\mu=0.16\pm 6.0\times 10^{-2}$\\ $\gamma=2.0\pm 3.4\times 10^{-1}$\end{tabular}  & \begin{tabular}[c]{@{}l@{}}$p:~ 0.21\pm 2.2\times 10^{-2}$\\ $q_{pmt}:~ 0.38\pm 1.9\times 10^{-2}$\end{tabular} & \begin{tabular}[c]{@{}l@{}}$p:~0.78\pm 9.6\times 10^{-2}$\\ $u:~1.4\pm 1.6\times 10^{-1}$\\ $q_{pmt}:~0.40\pm 1.9\times 10^{-2}$\end{tabular} \\ \cline{1-4}

\begin{tabular}[c]{@{}l@{}}$\alpha=0.2$\\ $\beta=1.7$\\ $\mu=0.2$\\ $\gamma=3.2$\end{tabular}                                               & \begin{tabular}[c]{@{}l@{}}$\alpha=0.2\pm 4.3\times 10^{-3}$\\ $\beta=1.7\pm 2.2\times 10^{-2}$\\ $\mu=0.18\pm 2.8\times 10^{-2}$\\$\gamma=3.5\pm 7.9\times 10^{-2}$\end{tabular}  & \begin{tabular}[c]{@{}l@{}}$p:~ 0.36\pm 9.2\times 10^{-2}$\\ $q_{pmt}:~ 0.11\pm 1.9\times 10^{-2}$\end{tabular} & \begin{tabular}[c]{@{}l@{}}$p:~0.46\pm 5.0\times 10^{-2}$\\ $u:~0.98\pm 3.4\times 10^{-2}$\\ $q_{pmt}:~0.15\pm 1.4\times 10^{-2}$\end{tabular} \\ \cline{1-4}

\begin{tabular}[c]{@{}l@{}}$\alpha=0.2$\\ $\beta=1.7$\\ $\mu=0.3$ \\$\gamma=3.2$\end{tabular}                                               & \begin{tabular}[c]{@{}l@{}}$\alpha=0.2\pm 6.0\times 10^{-3}$\\ $\beta=1.6\pm 1.4\times 10^{-1}$\\ $\mu=0.25\pm 7.5\times 10^{-2}$\\ $\gamma=3.5\pm 1.5\times 10^{-2}$\end{tabular}  & \begin{tabular}[c]{@{}l@{}}$p:~ 0.40\pm 1.2\times 10^{-2}$\\ $q_{pmt}:~ 0.19\pm 1.6\times 10^{-2}$\end{tabular} & \begin{tabular}[c]{@{}l@{}}$p:~0.57\pm 2.2\times 10^{-2}$\\ $u:~2.3\pm 2.0\times 10^{-1}$\\ $q_{pmt}:~0.23\pm 1.5\times 10^{-2}$\end{tabular} \\ \cline{1-4}
\end{tabular}\captionof{table}{\textbf{Train and test errors for case 4 in V-lock-in region (\S \ref{sec:inv_prob_syn}):} Table showing the actual parameters for the simulation (first column), the corresponding predicted values (second column), along with the train error in the data (third column) and test (fourth column) errors in the PINN solution. The frequency ratio $f=0.8$ corresponds to the V-lock-in region.}\label{tab:v_lock_in_num}

\begin{tabular}{llll}
\cline{1-4}
True value                                                                                                                                  & PINN predicted value                                                                                                                                                                          & Train error $E_r$ \%                                                                                                                                                                                            & Test error $E_r$ \%                                                                                                                                                                         \\ 
\cline{1-4}

\begin{tabular}[c]{@{}l@{}}$\alpha=0.4$\\ $\beta=1.3$\\ $\mu=0.3$\\ $\gamma=1.3$\end{tabular}                                               & \begin{tabular}[c]{@{}l@{}}$\alpha=0.39\pm 1.2\times 10^{-2}$\\ $\beta=1.3\pm 4.2\times 10^{-2}$\\ $\mu=0.37\pm 1.2\times 10^{-2}$\\ $\gamma=1.2\pm 1.6\times 10^{-2}$\end{tabular}  & \begin{tabular}[c]{@{}l@{}}$p:~ 0.12\pm 3.0\times 10^{-2}$\\ $q_{pmt}:~ 0.2\pm 1.4\times 10^{-2}$\\ $f=1.05$, case 4 \end{tabular} & \begin{tabular}[c]{@{}l@{}}$p:~0.20\pm 4.5\times 10^{-2}$\\ $u:~0.33\pm 1.9\times 10^{-2}$\\ $q_{pmt}:~0.23\pm 1.8\times 10^{-2}$\end{tabular} \\ \cline{1-4}
\begin{tabular}[c]{@{}l@{}}$\alpha=0.4$\\ $\beta=1.3$\\ $\mu=0.2$\\$\gamma=1.3$\end{tabular}                                              & \begin{tabular}[c]{@{}l@{}}$\alpha=0.36\pm 7.6\times 10^{-3}$\\ $\beta=1.2\pm 2.6\times 10^{-2}$\\ $\mu=0.21\pm 2.8\times 10^{-2}$\\ $\gamma=1.4\pm 1.1\times 10^{-2}$\end{tabular}  & \begin{tabular}[c]{@{}l@{}}$p:~ 0.29\pm 3.4\times 10^{-2}$\\ $q_{pmt}:~ 0.12\pm 1.4\times 10^{-2}$\\ $f=0.93$, case 4 \end{tabular} & \begin{tabular}[c]{@{}l@{}}$p:~0.36\pm 4.0\times 10^{-2}$\\ $u:~1.0\pm 1.7\times 10^{-1}$\\ $q_{pmt}:~0.14\pm 1.3\times 10^{-2}$\end{tabular} \\ \cline{1-4}
\begin{tabular}[c]{@{}l@{}}$\alpha=0.4$\\ $\beta=1.3$\\ $\mu=0.3$\\ $\gamma=1.3$\end{tabular}                                               & \begin{tabular}[c]{@{}l@{}}$\alpha=0.38\pm 9.1\times 10^{-3}$\\ $\beta=1.2\pm 3.9\times 10^{-2}$\\ $\mu=0.24\pm 2.4\times 10^{-2}$\\ $\gamma=1.4\pm 1.1\times 10^{-2}$\end{tabular}  & \begin{tabular}[c]{@{}l@{}}$p:~ 0.26\pm 4.9\times 10^{-2}$\\ $q_{pmt}:~ 0.10\pm 5.0\times 10^{-2}$\\ $f=0.9$, case 4 \end{tabular} & \begin{tabular}[c]{@{}l@{}}$p:~0.34\pm 6.8\times 10^{-2}$\\ $u:~1.1\pm 3.0\times 10^{-1}$\\ $q_{pmt}:~0.13\pm 4.7\times 10^{-2}$\end{tabular} \\ \cline{1-4}

\begin{tabular}[c]{@{}l@{}}$\alpha=0.4$\\ $\beta=1.3$\\ $\mu=0.3$\\ $\gamma=1.3$\end{tabular} &
\begin{tabular}[c]{@{}l@{}}$\alpha=0.38\pm 1.8\times 10^{-2}$\\ $\beta=1.1\pm 1.3\times 10^{-1}$\\ $\mu=0.31\pm 1.8\times 10^{-2}$\\ $\gamma=1.3\pm 8.0\times 10^{-1}$\end{tabular}  & \begin{tabular}[c]{@{}l@{}}$p:~ 0.26\pm 6.3\times 10^{-2}$\\ $q_{pmt}:~ 0.19\pm 6.3\times 10^{-2}$\\ $f=0.9$, case 4a \end{tabular} & \begin{tabular}[c]{@{}l@{}}$p:~0.45\pm 1.9\times 10^{-1}$\\ $u:~2.0\pm 9.6\times 10^{-1}$\\ $q_{pmt}:~0.21\pm 7.6\times 10^{-2}$\end{tabular} \\ \cline{1-4}
\end{tabular}\captionof{table}{\textbf{Train and test errors for case 4 in A-lock-in region (\S \ref{sec:inv_prob_syn}):} Table showing the actual LOM parameters for the simulation (first column), the corresponding predicted values (second column), along with the train error in the data (third column) and test error (fourth column) in the PINN solution. The value of the frequency ratio is provided in the third column, which corresponds to the A-lock-in region. The last two sets have the same parameters while differing in the loss-weights (case 4 \& 4a)}\label{tab:a_lock_in_num}
\end{center}
\section{Figures illustrating pointwise ensemble average and self-adaptive weight fields}
\begin{figure}
\includegraphics[keepaspectratio,width=1\textwidth]{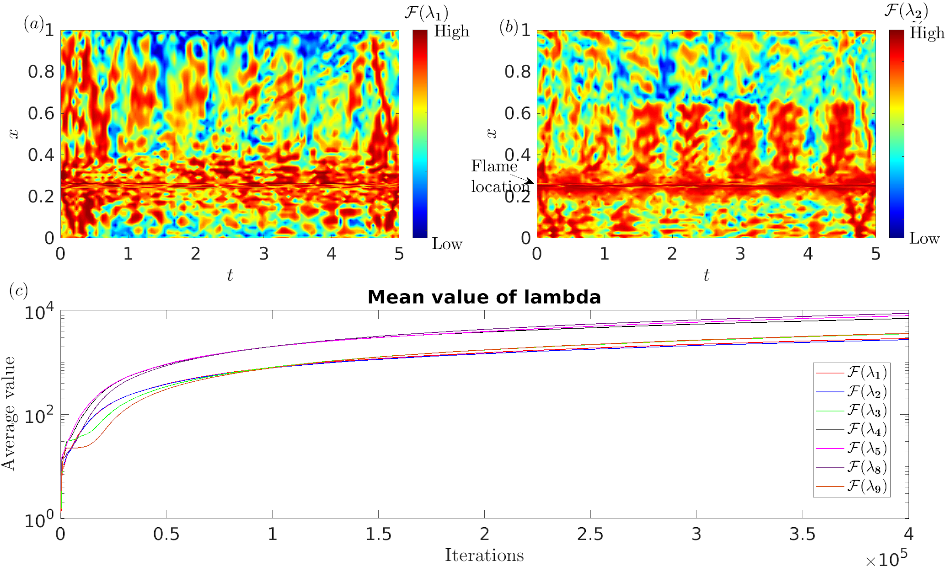}
\caption{\textbf{Mask function applied self-adaptive weights for case 4 (\S \ref{sec:inv_prob_syn}):} Self-adaptive weights with mask function at the end of the last iteration for acoustic pressure (panel $a$) and velocity (panel $b$) fields. The weights are high near the sharp gradient around the flame. The gradients in the acoustic wave pattern are also captured. Bottom panel $(c)$ indicates the average over the field of all the self-adaptive weights used in case 4. The first 400000 iterations are shown to highlight the initial iterates of the self-adaptive weights. The figure corresponds to the solution of figure \ref{fig:initial_condition_check_V_A_lock_i}($e,f$).}
\label{fig:self_adap_weights_case_7}
\end{figure}

\begin{figure}
\includegraphics[keepaspectratio,width=1\textwidth]{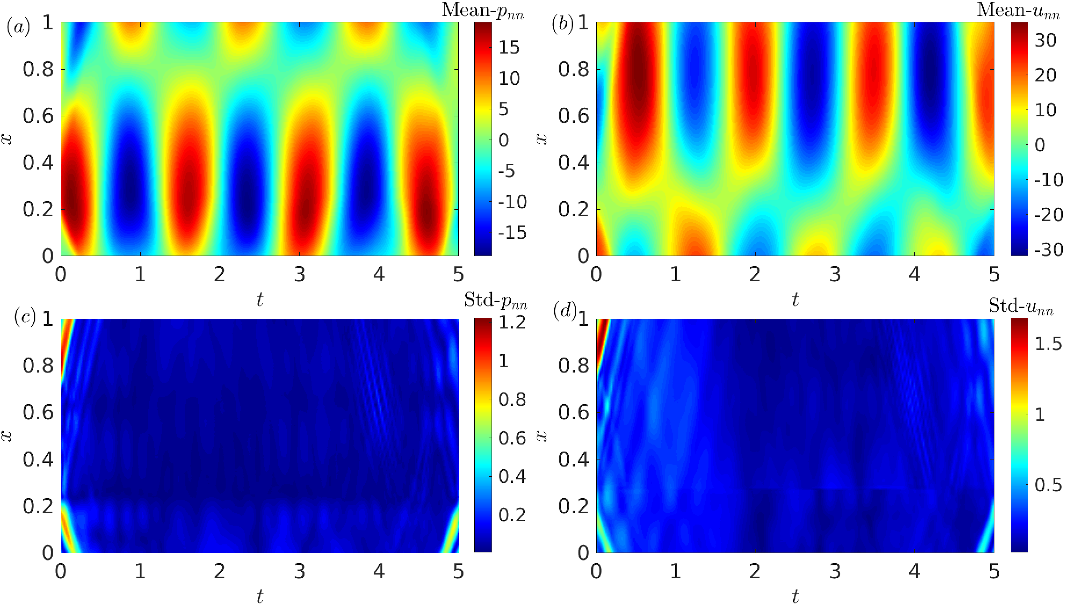}
\includegraphics[keepaspectratio,width=1\textwidth]{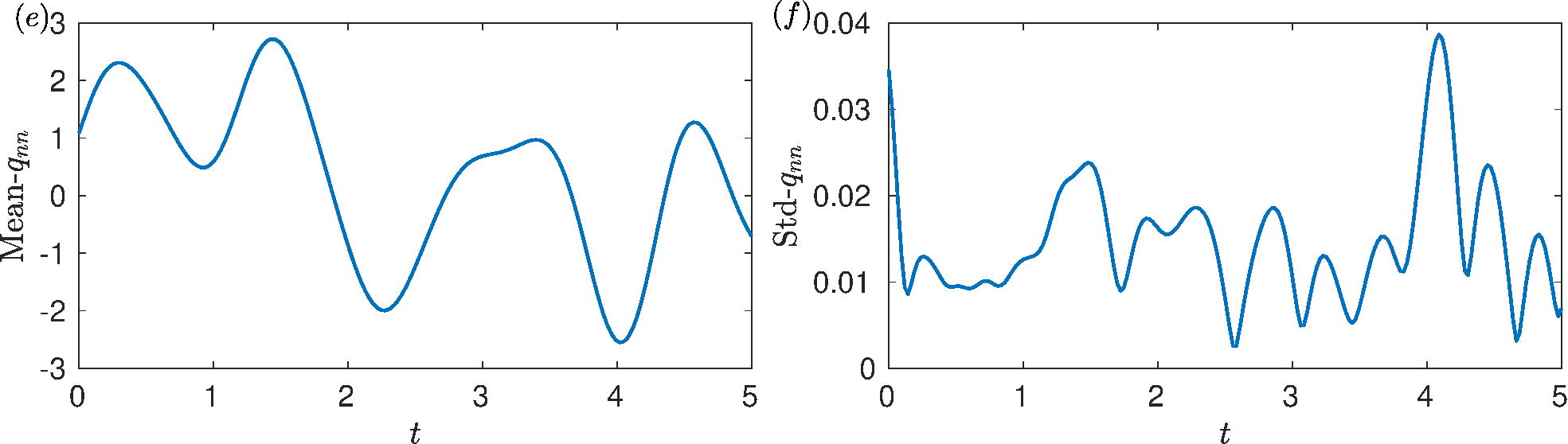}
\caption{\textbf{Ensemble mean and standard deviation of the PINN solution for case 5 (\S \ref{sec:inv_prob_exp}):} Mean $(a,~b,~e)$ and standard deviation $(c,~d,~f)$ of the PINN solution of acoustic pressure, velocity, and the unsteady heat release rate due to random initialization. The results correspond to the first chunk ($t=0-5$) of figure \ref{fig:42_slpm_stitching}.}
\label{fig:42_slpm_multiple_runs_std}
\end{figure}

\end{document}